\documentclass[aps
amsmath, amssymb, twocolumn,superscriptaddress, floats, showkeys, showpacs]
{revtex4-2}
\usepackage{physics}
\usepackage{float}
\usepackage[caption=false]{subfig}
\usepackage{ulem}
\usepackage[euler]{textgreek}
\usepackage{mathrsfs}
\usepackage{graphicx}
\usepackage{natbib}
\usepackage{amsmath}
\usepackage{textcomp}
\usepackage{epstopdf}
\usepackage[dvipsnames]{xcolor}
\definecolor{MMgreen}{RGB}{0,128,0}
\usepackage{sidecap}
\newcommand{\unit}[1]{\ensuremath{\, \mathrm{#1}}}
\renewcommand{\vec}[1]{\mathbf{#1}}
\usepackage{xr-hyper}
\usepackage{hyperref}
\hypersetup{colorlinks,citecolor=blue, filecolor=blue, linkcolor=blue , urlcolor=blue}
\newcommand{\comment}[1]{}

\begin{document}

\title{Exchange-split multiple Rydberg series of excitons \\
in anisotropic quasi two-dimensional ReS$_{2}$}%

\author{P.~Kapu\'sci\'nski}
\email{piotr.kapuscinski@lncmi.cnrs.fr}
\affiliation{LNCMI-EMFL, CNRS UPR3228, Univ. Grenoble Alpes, Univ. Toulouse, Univ. Toulouse 3, INSA-T,  
Grenoble and Toulouse, France}
\affiliation{Department of Experimental Physics,
Wrocław University of Science and Technology, 50-370
Wrocław, Poland}
\author{J.~Dzian}
\affiliation{Department of Experimental Physics, Faculty of Science, Palacky University, 17. listopadu 17, 771 46 Olomouc, Czech Republic}
\author{A.~O.~Slobodeniuk}
\affiliation{Department of Condensed Matter Physics, Faculty of Mathematics and Physics,
Charles University, Ke Karlovu 5, CZ-121 16 Prague, Czech Republic}
\author{C.~Rodr\'iguez-Fern\'andez}
\affiliation{LNCMI-EMFL, CNRS UPR3228, Univ. Grenoble Alpes, Univ. Toulouse, Univ. Toulouse 3, INSA-T,  
Grenoble and Toulouse, France}
\author{J.~Jadczak}
\affiliation{Department of Experimental Physics,
Wrocław University of Science and Technology, 50-370
Wrocław, Poland}
\author{L.~Bryja}
\affiliation{Department of Experimental Physics,
Wrocław University of Science and Technology, 50-370
Wrocław, Poland}
\author{C.~Faugeras}
\affiliation{LNCMI-EMFL, CNRS UPR3228, Univ. Grenoble Alpes, Univ. Toulouse, Univ. Toulouse 3, INSA-T,  
Grenoble and Toulouse, France}
\author{D.~M.~Basko}
\email{denis.basko@lpmmc.cnrs.fr}
\affiliation{Univ. Grenoble Alpes, CNRS, LPMMC, 38000 Grenoble, France}
\author{M.~Potemski}
\email{marek.potemski@lncmi.cnrs.fr}
\affiliation{LNCMI-EMFL, CNRS UPR3228, Univ. Grenoble Alpes, Univ. Toulouse, Univ. Toulouse 3, INSA-T,  
Grenoble and Toulouse, France}
\affiliation{Institute of Experimental Physics, Faculty of Physics, University of Warsaw, ul. Pasteura 5, 02-093 Warszawa, Poland}

\begin{abstract}
We perform a polarization-resolved magnetoluminescence study of excitons in ReS$_2$.
We observe that two linearly polarized Rydberg series of excitons are accompanied by two other Rydberg series of dark excitons, brightened by an in-plane magnetic field.
All series extrapolate to the same single-electron bandgap, indicating that the observed excitons originate either from the same valley or from two valleys related by the inversion symmetry, and are split by exchange interaction. 
To interpret our observations of the magnetic brightening, we have to assume the dominant spin-orbit coupling to be Ising-like, which hints at an approximate symmetry of the electronic states in ReS$_2$ which is higher than the crystal symmetry $C_i$. 
\end{abstract}

\maketitle

\section{Introduction \label{sec:intro}}

While the extensive experimental and theoretical works on group VI (Mo- and W-based) semiconducting transition metal dichalcogenides (S-TMDs) provided a fair understanding of basic, electronic and optical properties of these materials, their Re-based (group VII) counterparts are by far less understood. Even though all S-TMDs are layered compounds, with a single layer consisting of the transition metal atoms plane sandwiched between two chalcogen atom planes \citep{Chhowalla2013}, the highly contrasting in-plane symmetries impose a stark difference in the properties of group VI and group VII dichalcogenides. The group VI S-TMDs crystallize in $2H$ and $3R$ phases, with a hexagonal in-plane structure, making them nearly isotropic in the layers’ plane.  On the other hand, in the group VII S-TMDs, the strong interactions between rhenium atoms lead to the Jahn-Teller distortion, inducing the $1T'$ crystal phase with chains of Re$_4$ clusters \cite{Kertesz1984, LAMFERS199634, Murray1994}. This results in a robust in-plane anisotropy which, in particular, manifests itself in an optical response being sensitive to the direction of linear polarization of light  \cite{PhysRevB.58.16130, lin2011, HO2007245}. 

The strength of interlayer coupling is another parameter that differentiates the dichalcogenides of groups VI and VII. Layers are weakly coupled in all S-TMDs, but the interlayer coupling is still pronounced in group VI S-TMDs. The important evolution of the band structure in Mo- and W-dichalcogenides is observed as a function of the number of layers: from indirect- to the direct-gap semiconductors in the limit of single layers\cite{splendiani2010, mak2010, Ellis2011}. The latter are thus of particular interest, especially for their unique optical properties. On the other hand, the interlayer coupling seems to be much weaker in Re-dichalcogenides\cite{Tongay2014, Yu2015, Jariwala2016, Echeverry2018}. Consequently, their electronic and optical properties remain defined mainly by the intra-layer coupling parameters even in the limit of bulk crystals. Strong in-plane anisotropy, in combination with an optional advantage of working with bulk or multilayer structures, holds the promise to use Re-dichalcogenides in polarization-sensitive devices\cite{Sim2019, WangJunyong2020, ZhangEnze2015, Kwon2019}.

The optical response of Re-dichalcogenides is known to be governed by two distinct excitonic resonances, split in energy and linearly polarized along different in-plane directions \cite{Aslan2016, Jadczak2019SR, Urban_2018, Sim2018, Xiaofan2019, HO2019641,kipczak2020optical, Dhara2020AdditionalEF}. The origin of this fundamental property of ReSe$_2$ and ReS$_2$ is, however, ambiguous. The contradictory scenarios invoking the exchange mediated splitting (in ReSe$_2$)\cite{Arora2017}, the transitions at two different, Z and K1 points of the Brillouin zone \cite{Dhara2020AdditionalEF} or involving different orbitals have been proposed\cite{HO2019641}. Besides, no clear consensus has been reached concerning even such a basic property as the direct or indirect character of the fundamental bandgap in Re-dichalcogenides. Recent experiments point towards an indirect gap in bulk and thin layer structures \cite{Urban_2018, Xiaofan2019,Guti_rrez_Lezama_2016,Ho_1998,Ho_1999, Ho_2004,Zelewski2017}, whereas theoretical calculations are not conclusive on this issue \cite{Echeverry2018,Gunasekera2018}.

\begin{figure*}[bt]
	\includegraphics[width=14.5cm]{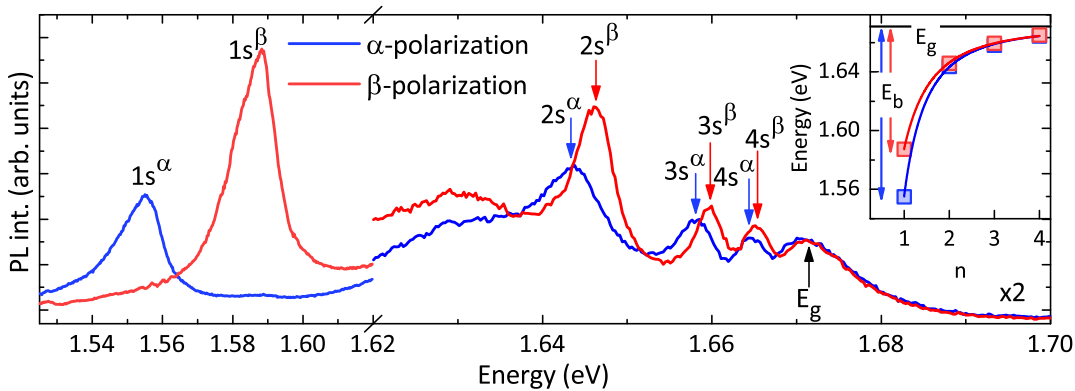}
	\caption{$\alpha$-polarized (blue) and $\beta$-polarized (red) low-temperature photoluminescence response of the bulk ReS$_2$ sample. In the energy region of the excited transitions, the energy scale has been stretched, and the spectra were multiplied by a factor of 2 for clarity. The inset presents the experimentally obtained transition energies for the ns$^{\alpha}$ and ns$^{\beta}$ exciton states as a function of principal quantum number $n$. Solid blue and red lines show fits to the data by Eq.~(\ref{eq:1}). The black vertical arrow and horizontal line on the main figure and the inset, respectively, highlight the bandgap energy $E_g$ obtained from the fits, while blue and red double-headed arrows on the inset represent corresponding binding energies $E_b$.}
	\label{fig:0T}
\end{figure*}

In the present work we elucidate the nature of excitonic resonances in bulk ReS$_2$ crystals. On the experimental ground this is done with high-field magneto-luminescence measurements, with which we investigate the linearly polarized emission spectra in different configurations of the direction of the applied field with respect to the crystal axes. The experimental observations are confronted with the theoretical model for interband transitions in ReS$_2$, which respects the appropriate $C_i$ symmetry of this crystal. Central to our work is the investigation of the Rydberg excitonic series, with the ground and excited states, and the original observation of two series of ``dark'' excitons in complement to the commonly observed two other series of ``bright'' excitons. All four excitonic series extrapolate to the same single-particle band gap, pointing at their origin either at a single high-symmetry point of the Brillouin zone, or at two different valleys, related by the space inversion. The presence of two bright and two dark excitons which are mixed by the in-plane magnetic field suggests that the dominant spin-orbit coupling is Ising-like; this implies that the crystal has an approximate symmetry that is higher than the known~$C_i$. Notably, our experiments do not exclude the possibility of ReS$_2$ being an indirect semiconductor, but they indicate that direct and indirect gaps must be relatively close in energy.

\section{Results \label{sec:results}}

\subsection{Emission resonance in the absence of the magnetic field \label{results:resonances}}

The low-temperature photoluminescence (PL) response of bulk ReS$_2$  is composed of several distinct resonances. As already shown in previous reports \cite{Aslan2016, Urban_2018, Sim2018, Jadczak2019SR, HO2019641, Xiaofan2019}, the two lowest in energy transitions are almost perfectly linearly polarized. 
When rotating the axis of the linear polarizer placed in the collection beam, one can extinguish either one or the other emission peak. This occurs at two distinct polarization directions, and we refer to $\alpha$- and $\beta$- polarization configurations for spectra with dominant lower-energy and higher-energy peaks, respectively. 
For more details on unpolarized spectrum and its polarization dependence, see Appendix A and Fig.~\ref{fig:unp0T}. The $\alpha$- and $\beta$-polarized PL spectra  obtained at $T=4.2$ K are presented in Fig.~\ref{fig:0T}. This figure shows that $\alpha$- and $\beta$-polarized spectra are composed of two distinct sets of emission peaks. Following Ref.~\cite{Jadczak2019SR}, the lowest peak in each polarization is attributed to the ground states excitonic resonances whereas higher energy peaks correspond to the excited excitonic states. This can be particularly well seen when analyzing the energy diagrams of each series using the following expression:
\begin{equation} \label{eq:1}
E_{n}=E_{g}-\dfrac{\mathrm{Ry}^*}{(n+\gamma)^2},
\end{equation}
where $E_{n}$ is the energy of the $n$th peak ($n=1,2,\ldots$) in each series, $E_g$ is the single-particle bandgap associated with the corresponding series, $\mathrm{Ry}^*$ is the effective Rydberg constant, and $\gamma$ is a phenomenological fit parameter. As shown in the inset to Fig.~\ref{fig:0T}, the above formula fully reproduces the experimental data with parameters summarized in Table~\ref{tab:1}. 

\begingroup
\def\arraystretch{1.5}
\begin{table}[]
\caption{Parameters obtained by fitting Eq.~(\ref{eq:1}) to the energies of the $\alpha$- and $\beta$-polarized excitonic series. The experimental transition energies described by these parameters are depicted on the inset of Fig.~\ref{fig:0T}.
}
\begin{tabular*}{\columnwidth}{@{\extracolsep{\fill}}ccccc@{}}
\hline\hline
   & $E_g$ (eV) & $\mathrm{Ry}^*$ (meV) & $\gamma$ & $E_b$ (meV)  \\ \hline
$\alpha$ & 1.671   & 110       & -0.03  & 117   \\
$\beta$ & 1.671   & 123       & 0.21 & 84      \\
\hline\hline
\end{tabular*}

	\label{tab:1}
\end{table}
\endgroup

Notably, both  series extrapolate to the same bandgap. As can be observed from the inset of Fig.~\ref{fig:0T}, the energy of this bandgap falls in the spectral range of a relatively broad emission band with the characteristic high energy tail. This feature, terminating the measured PL spectra on the high energy side, is therefore attributed to the expected response due to a continuum of single-particle electron-hole excitations. Emission peaks due to excitonic states with $n>4$ are not resolved but are expected to be spread within the high energy part of the $n=4$ emission peak.

\begin{figure*}[bt]
	\includegraphics[width=14.5cm]{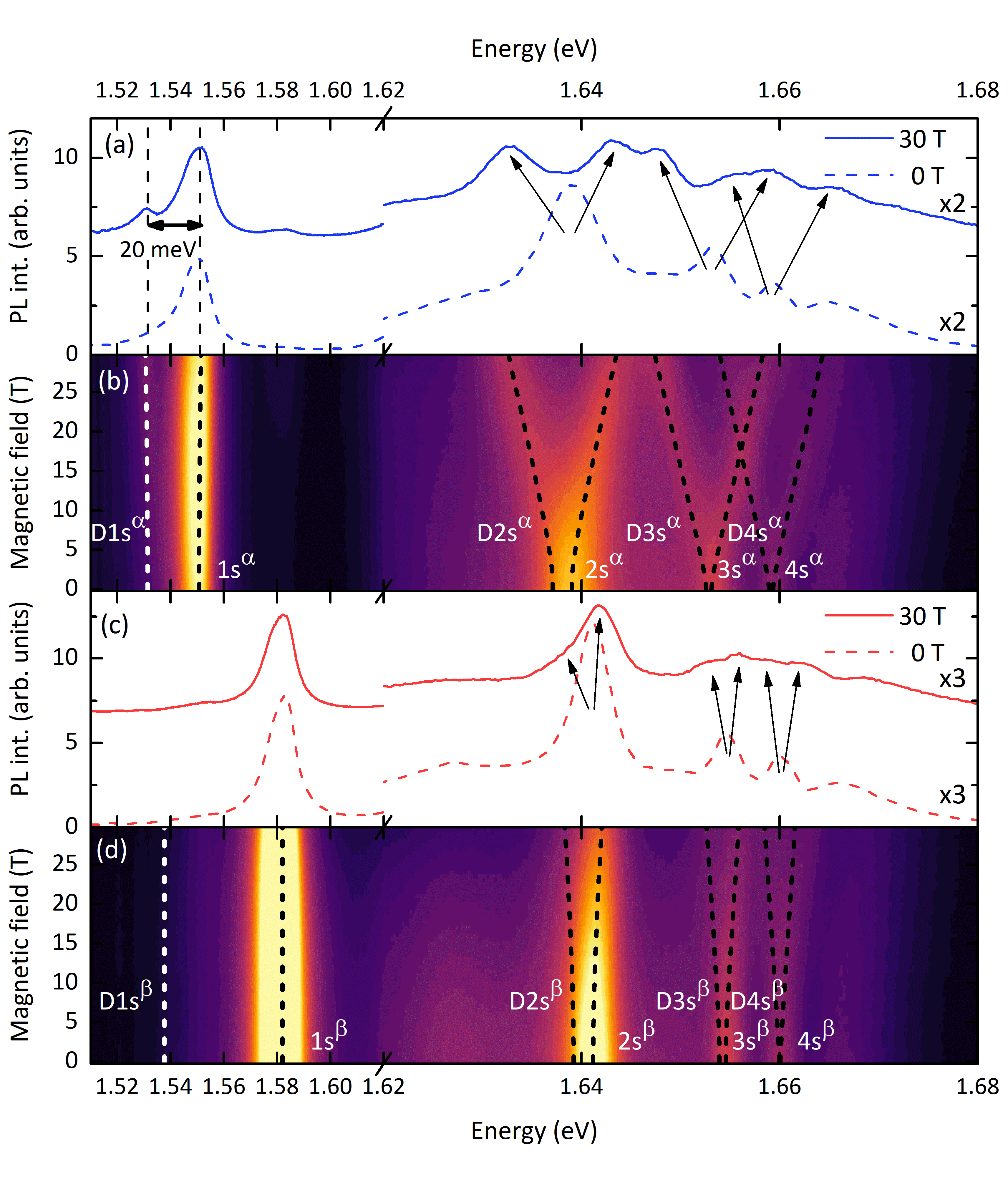}
	\caption{The results of magneto-photoluminescence measurements performed in the Voigt geometry with $B\parallel b$. Comparison of the spectra obtained for (a) $\alpha$- and (c) $\beta$-linear polarization at $B=0$~T (dashed lines) and $B=30$~T (solid lines). The spectra has been vertically shifted for clarity. The black arrows follow the apparent splitting of the excited states. False-colour maps of the PL response as a function of magnetic field for (b) $\alpha$- and (d) $\beta$-linear polarizations. Dashed vertical white and black lines represent transition energies obtained from the theoretical model. The energy scales on all figures have been stretched in the energy region of the excited transitions. }
	\label{fig:Voi1}
\end{figure*}

Equation (\ref{eq:1}) is well known for bound states in the isotropic hydrogenic problem with the dimensionality $d=2,3$, where the corresponding $\gamma=-1/2,0$, respectively.
It also was also derived for a non-integer dimensionality $d$ of the space, where $\gamma=(d-3)/2$ \cite{He1991,Mathieu1992,PhysRevB.46.4092}. This expression for fractional~$d$ was found to successfully describe the experimentally measured energies for excitons in quantum wells~\cite{Christol1993} and TMDs~\cite{Molas5s} with in-plane isotropy and non-hydrogenic potential. With in-plane isotropy, excitonic states can be classified according to the relative angular momentum ($s$, $p$, $d$, etc.), and only $s$~excitons are optically active. In ReS$_2$, there is no in-plane isotropy, so $s$ states can be coupled to $d$ states, as well as higher even angular momenta (note that the parity of the angular momentum is preserved by the inversion symmetry of the two-particle Schr\"odinger equation, which holds even in a low-symmetry valley, when the electron and hole dispersion are expanded to the quadratic order in momentum). Thus, strictly speaking, all excitonic states with even parity of the relative electron-hole motion are expected to be optically active. This would lead to several series of peaks in each polarization, which we do not observe. We thus conclude that the observed series in each polarization corresponds to states with the dominant $s$~wave character, while those dominated by higher angular momenta, even though allowed by symmetry, have too small oscillator strength to be resolved in our experiment. Thus, we label the observed excitonic peaks as $ns^\alpha$ and $ns^\beta$ with $n=1,2,3,4$.

\subsection{Magnetophotoluminescence in Voigt geometry \label{results:voigt}}

\begin{figure*}[tb]
	\includegraphics[width=14.5cm]{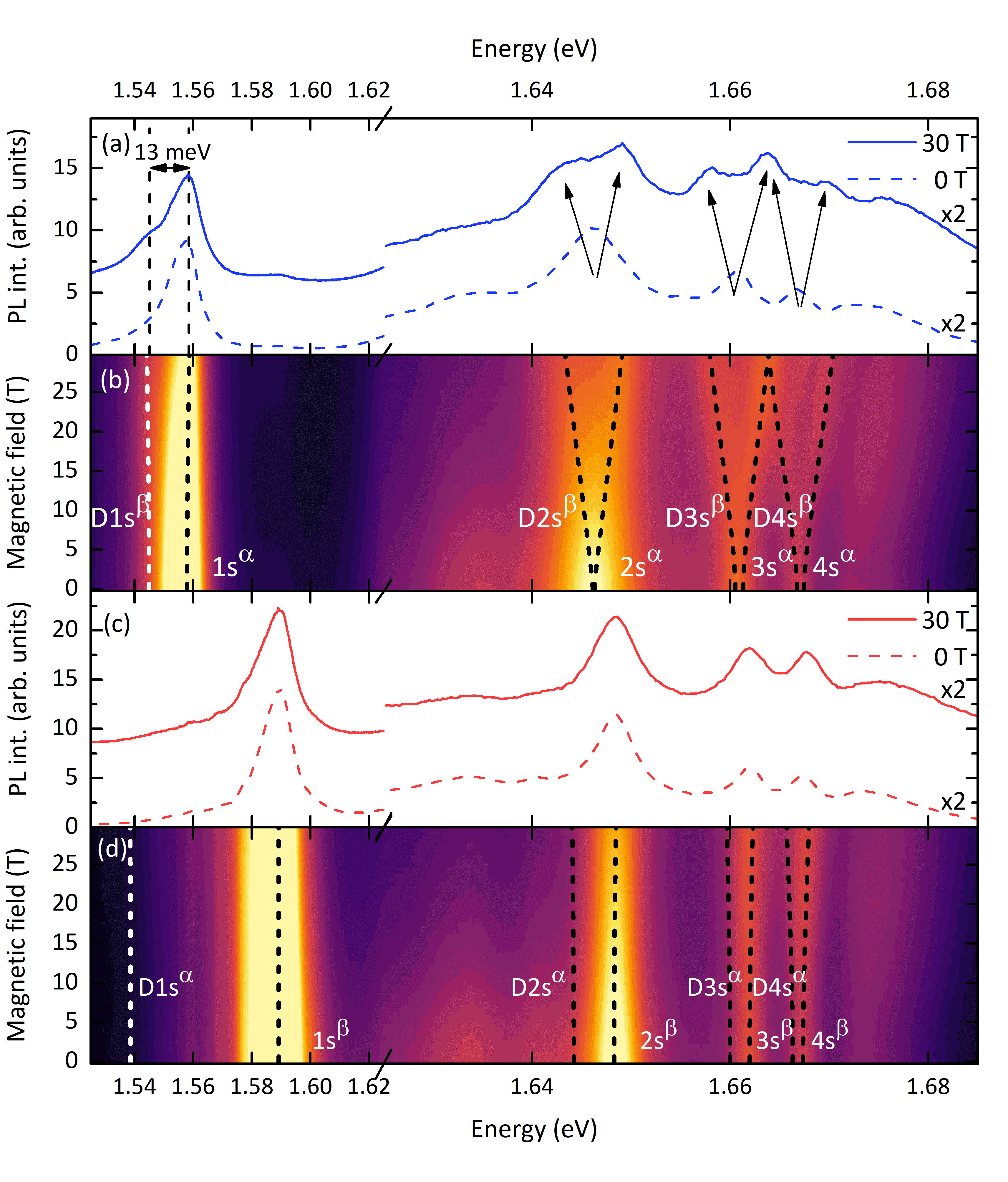}
	\caption{The results of magneto-photoluminescence measurements performed in the Voigt geometry with $B\perp b$. Comparison of the spectra obtained for (a) $\alpha$- and (c) $\beta$-linear polarization at $B=0$~T (dashed lines) and $B=30$~T (solid lines). The spectra has been vertically shifted for clarity. The black arrows follow the apparent splitting of the excited states. False-colour maps of the PL response as a function of magnetic field for (b) $\alpha$- and (d) $\beta$- linear polarizations. Dashed vertical white and black lines represent transition energies obtained from the theoretical model. The energy scales on all figures have been stretched in the energy region of the excited transitions.  Note that the absolute energies of the resonances differ slightly from the measurements presented in Fig.~\ref{fig:Voi1}. Such differences are observed from sample to sample; the relative energies, however, do not change, see Fig.~\ref{fig:unp0T}(b) in Appendix A. }
	\label{fig:Voi2}
\end{figure*}

To get more insight about the fine structure of excitons observed in bulk ReS$_2$, we turn to magnetophotoluminescence experiments with the magnetic field $B$ applied in Voigt geometry, which is a technique widely used to reveal otherwise optically inactive (dark) excitons in S-TMDs \cite{molas, wang2017, Zhang2017, Lu_2019, Robert2020MeasurementOT, zinkiewicz2020neutral}. Here, due to the optical anisotropy, both the linear polarization of the collected light and the direction of the in-plane field with respect to the crystallographic axis should be controlled separately. For simplicity, we use only two previously introduced $\alpha$ and $\beta$ linear polarization configurations: the magnetic field is applied either along or perpendicularly to the rhenium chains (which in our convention defines the direction of $b$ crystal axis). With reference to similar experiments carried out on W- or Mo-based dichalcogenides \cite{molas, wang2017, Zhang2017, Lu_2019, Robert2020MeasurementOT, zinkiewicz2020neutral} and also on anisotropic III/V and II/VI quantum dots \cite{Bayer2002, Kowalik2007}, one may expected a “magnetic brightening” of certain transitions, which are “dark” in the absence of the magnetic field. Indeed such effect is observed in our experiments: the magnetically brighten resonances are denoted as D$ns^{\alpha}$ and D$ns^{\beta}$ emission peaks.

\subsubsection{$B\parallel b$ \label{results:voigt1}}

Fig.~\ref{fig:Voi1} summarizes the results of the magnetophotoluminescence measurements in Voigt geometry with $B$ parallel to the rhenium chains in magnetic fields up to 30 T for $\alpha$ and $\beta$ configurations of linear polarizations. Two main effects can be noted when the magnetic field is applied. Firstly, the additional resonance appears in the spectrum in the $\alpha$ polarization, $\approx 20$~meV below $1s^{\alpha}$ transition (see Fig.~\ref{fig:Voi1}(a)), which we assign here as D1s$^{\alpha}$.  
No similar effect is observed in the $\beta$ polarization configuration.

Secondly, all the excited states of the $ns^{\alpha}$ series are seemingly split into two lines (see Fig.~\ref{fig:Voi1}(a) and (b)). In a first approximation, the magnitude of this splitting is the same for all states. Excited states of the $ns^{\beta}$ series undergo a similar splitting, but more than 3 times smaller in magnitude (see Fig.~\ref{fig:Voi1}(c) and (d)).

Both effects can be explained by the existence of the dark exciton series D$ns^{\alpha}$, whose ground state lies $\approx 20$~meV below the $ns^{\alpha}$ ground state, while the excited states are close in energy to the ones of the bright series. When a magnetic field is applied, dark and bright states are  mixed, leading to the brightening of dark states, as observed for the ground dark state D1s$^{\alpha}$. Additionally, if they are close to each other in energy, as it is the case of the excited states,  they seem to repel each other (as in the case of a MoSe$_2$ monolayer\cite{Lu_2019, Robert2020MeasurementOT}), likely due to an in-plane Zeeman term. These two effects lead to an apparent splitting of the excited bright states.

\subsubsection{$B\perp b$ \label{results:voigt2}}

Similar effects are observed when the magnetic field is applied perpendicularly to the $b$-axis, see Fig.~\ref{fig:Voi2}. However, as can be seen in  Fig.~\ref{fig:Voi2}(a), the  energy of the brightened D1s$^{\beta}$ state is only of about $\approx 13$~meV lower that that of 1s$^{\alpha}$ state.
This implies that the observed D1s$^{\beta}$ state is associated to the second, $\beta$-series of dark excitons. 

In this configuration, the ``splitting'' of the excited states of  $ns^{\alpha}$ series is notably weaker with magnitude roughly twice smaller than in $B\parallel b$ configuration. No such effect can be measured for the $ns^{\beta}$ series (Fig.~\ref{fig:Voi2}(c) and (d)) - only small broadening and blueshift of the excited lines is observed.

\subsubsection{Recap of experimental observations}

The experimental results presented so far can be summarized in the following way: the linear-polarization-resolved PL experiments in a zero-magnetic field showed the presence of two non-degenerate Rydberg series of excitons that extrapolate to the same bandgap. The angle between linear-polarizations ($\alpha$- and $\beta$-polarizations) of these two series is found to be 78$^\circ$. The magneto-PL measurements in two in-plane field geometries yield a clear demonstration of the brightening effect of two non-degenerate dark excitons' ground states, but only observed in $\alpha$-polarization configuration. The brightening of the ground states is accompanied by the apparent splitting of excited states of both $\alpha$- and $\beta$-polarized bright excitons, which can be accounted by the repulsion of interacting bright and dark excited states. However, the strength of this repulsion depends on the direction of the field and is notably weaker in the $\beta$-polarization.  Similar measurements performed in Faraday geometry did not show any signs of  brightening of dark excitons (see Appendix B).

\subsection{Theoretical background \label{sec:phenomodel}}

The goal of this section is to propose a theoretical model for excitons in ReS$_2$, based on the $C_i$ point symmetry group of the crystal (which is preserved in the presence of a magnetic field). The existence of dark excitons which can be brightened by a magnetic field is natural to interpret in terms of spin-related selection rules. Indeed, transitions forbidden by momentum or parity conservation could not be brightened by a magnetic field.

As shown above, the excited states of all observed excitonic species extrapolate to the same electronic band gap, so the excitons must originate either from a single valley in a high-symmetry point of the electronic first Brillouin zone, or from two valleys related to two inequivalent points $\pm\vec{k}$, related by the space inversion and the time reversal symmetry. These require that in each valley the electronic states are doubly degenerate, since the space inversion flips the momentum but not the spin, while the time reversal flips both momentum and spin. We study the one-valley and two-valley scenarios separately in the Appendix C and conclude that the available experimental data do not allow one to distinguish between these two cases. Namely, in the two-valley case, the odd-parity linear combinations of zero-momentum excitonic states from the two valleys behave quite analogously to the excitons in the single-valley case.

Let us see what constraints on the possible theoretical model are imposed by the experimental observations, reported above. Observations relevant to our considerations include:
(i)~there are two bright exciton states, visible in two different non-orthogonal linear polarizations, labelled $\alpha$ and $\beta$, with energy splitting of 35~meV ($\alpha$~referring to the low-energy component);
(ii)~there are two dark exciton states with energy splitting of 7~meV; 
(iii)~an in-plane magnetic field along the $b$~axis couples the low-energy dark exciton to the bright $\alpha$~exciton, and weakly couples the high-energy dark exciton to the bright $\beta$ exciton;
(iv)~an in-plane magnetic field  perpendicular to the $b$~axis couples the high-energy dark exciton to the bright $\alpha$ exciton.

Existence of two dark states out of four at zero field imposes a certain restriction on the spin-orbit interaction (SOI). If SOI is absent, then an optical transition conserves both the projection $S^z$ on the $z$~axis and the magnitude $S$ of the total spin $\vec{S}=\vec{s}_\mathrm{e}+\vec{s}_\mathrm{h}$ of the electron and the hole. Then, only the singlet state ($S=0$) can be bright, while three triplet states ($S=1$) must be dark, which contradicts the experimental observation. On the other hand, a strong SOI of a general form compatible with the inversion symmetry would make all four excitons bright at zero field. In order to have two bright and two dark states, one has to assume SOI of a special form, which conserves the projection of $\vec{S}$ on some axis (which we will call~$z$), but not~$S$, such as the Ising-type SOI occurring in hexagonal TMDs.

Labelling the electronic states in the conduction/valence band by their $z$ spin projection $s_\mathrm{c/v}=\uparrow,\downarrow$, we label the four exciton spin states as $|s_\mathrm{c},s_\mathrm{v}\rangle$. These are split by the exchange interaction whose structure is analysed in detail in the Appendix C. Crucially, the same $S^z$ conservation that forbids the optical transitions to the dark exciton states, also requires the splitting of the dark excitons to vanish. This contradicts the experimental observation of a finite (7~meV) splitting between the two dark exciton states, although significantly weaker than the splitting between the bright excitons (35~meV). To reconcile the two experimental observations, we can only speculate that the SOI, in addition to the main contribution which conserves $S^z$ but not~$S$, has a weaker contribution which breaks $S^z$ conservation, as allowed by the $C_i$ symmetry of the crystal. This contribution must induce a weak exchange splitting of the dark states and a weak brightening of the dark states, however, we cannot reliably estimate the relative magnitude of the two effects; there may be some small factor of purely numerical origin that make the SOI-induced brightening unobservable in our experiment.

Thus, we write the Hamiltonian of the four exciton states at zero magnetic field in the basis $|\uparrow_\mathrm{c}\uparrow_\mathrm{v}\rangle$, 
$|\downarrow_\mathrm{c}\downarrow_\mathrm{v}\rangle$,
$|\uparrow_\mathrm{c}\downarrow_\mathrm{v}\rangle$, 
$|\downarrow_\mathrm{c}\uparrow_\mathrm{v}\rangle$
as follows:
\begin{equation}\label{eq:exchange_Hamiltonian}
H_\mathrm{ex}=\left[\begin{array}{cccc}
b_0 & b_1 & 0 & 0 \\ b_1^* & b_0 & 0 & 0 \\
0 & 0 & d_0 & d_1 \\ 0 & 0 & d_1^* & d_0
\end{array}\right].
\end{equation}
Here $b_0$ and $b_1$ are the exchange induced shift and coupling of the bright excitons, discussed in detail in the Appendix C. To account for the experimentally observed splitting of the dark excitons, we introduce by hands analogous matrix elements $d_0,d_1$ in the dark sector. An $S^z$-non-conserving SOI should also produce non-zero matrix elements in the dark-bright and bright-dark blocks; however, we put them to zero (again, by hands) since we want to describe the only experimentally observed brightening mechanism, due to the in-plane magnetic field.

In the presence of a magnetic field~$\vec{B}$, we include its Zeeman coupling to the electron spin in the conduction/valence band:
\begin{equation}
H_Z^\mathrm{c/v}=\mu_B\sum_{k=x,y,z}\sum_{\kappa=\xi,\eta,\zeta}B_{\kappa}g_{\kappa{k}}^\mathrm{c/v}s^k_\mathrm{c/v}.
\end{equation}
Any real matrices $g_{\kappa{k}}^\mathrm{c},g_{\kappa{k}}^\mathrm{v}$ are allowed by the crystal symmetry~$C_i$ since both magnetic field and spin are invariant under inversion. Moreover, we have to introduce two different coordinate systems ($x,y,z$) and ($\xi,\eta,\zeta$), since we have no {\it a priori} knowledge how the $z$~direction (the one along which the spin projection is conserved) is related to the $\zeta$ direction (which we define to be the one perpendicular to the layers). The operators of the three electron spin components along the $x,y,z$ directions are determined by the corresponding Pauli matrices: $s^{x,y,z}=\sigma^{x,y,z}/2$. The $\eta$ direction will be associated with the $b$ crystalline axis.
Then the fact that a magnetic field perpendicular to the plane does not brighten the dark excitons (see Appendix B) implies $g_{\zeta{x}}^\mathrm{c/v}=g_{\zeta{y}}^\mathrm{c/v}=0$.

The exchange matrix elements $b_1$ and $d_1$ can be made real and positive by adjusting the phases of the wave functions in the conduction and valence bands, as discussed in detail in the Appendix C.
Then the bright and dark blocks of the Hamiltonian~(\ref{eq:exchange_Hamiltonian}) can be diagonalised by passing to symmetric and antisymmetric linear combinations. Omitting the Zeeman terms proportional to $s_\mathrm{c}^z,s_\mathrm{v}^z$ which do not couple the dark and the bright sector, we write the rotated Hamiltonian as
\begin{widetext}
\begin{equation}\label{eq:rotated_Hamiltonian}
\widetilde{H}
=\left[\begin{array}{cccc}
b_0+b_1 & 0 & 
\mathrm{Re}({\Omega}_\mathrm{c}-{\Omega}_\mathrm{v}) 
& i\mathrm{Im}({\Omega}_\mathrm{c}+{\Omega}_\mathrm{v}) \\
0 & b_0-b_1 & -i\mathrm{Im}({\Omega}_\mathrm{c}-{\Omega}_\mathrm{v}) & 
-\mathrm{Re}({\Omega}_\mathrm{c}+{\Omega}_\mathrm{v})  \\
\mathrm{Re}({\Omega}_\mathrm{c}-{\Omega}_\mathrm{v}) & 
i\mathrm{Im}({\Omega}_\mathrm{c}-{\Omega}_\mathrm{v}) & 
d_0+d_1 & 0  \\
-i\mathrm{Im}({\Omega}_\mathrm{c}+{\Omega}_\mathrm{v}) 
& -\mathrm{Re}(\Omega_\mathrm{c}+\Omega_\mathrm{v}) 
& 0  & d_0-d_1
\end{array}\right],
\end{equation}
\end{widetext}
where we denoted by
\begin{equation}
\Omega_\mathrm{c/v}=\frac{\mu_B}2\sum_{\kappa=\xi,\eta}B_{\kappa}(g_{\kappa{x}}^\mathrm{c/v}+ig_{\kappa{y}}^\mathrm{c/v}).
\end{equation}
the Zeeman matrix elements responsible for the brightening of the dark states.

The experimental observations (iii) and (iv) allow us to impose further restrictions on the model. The brightening of the two dark excitons for the two orthogonal magnetic field directions implies $\mathrm{Im}({\Omega}_\mathrm{c}-{\Omega}_\mathrm{v})=0$ and $\mathrm{Im}({\Omega}_\mathrm{c}+{\Omega}_\mathrm{v})=0$ for the field along~$b$ as well as $\mathrm{Re}({\Omega}_\mathrm{c}+{\Omega}_\mathrm{v})=0$ and $\mathrm{Re}({\Omega}_\mathrm{c}-{\Omega}_\mathrm{v})=0$ for the field perpendicular to~$b$. These conditions yield $g_{\eta y}^\text{c}=g_{\eta y}^\text{v}=0$ and $g_{\xi x}^\text{c}=g_{\xi x}^\text{v}=0$, respectively. Combining both results we obtain the Hamiltonian: 
\begin{equation}\label{eq:simpl_Hamiltonian}
\widetilde{H}=\left[\begin{array}{cccc}
b_0+b_1 & 0 &  \beta\mu_B B_\eta  & i\delta\mu_B B_\xi \\
0 & b_0-b_1 &  -i\gamma \mu_B B_\xi  & -\alpha\mu_B B_\eta \\
\beta\mu_B B_\eta & i\gamma\mu_B B_\xi & d_0+d_1 & 0  \\
-i\delta\mu_B B_\xi & -\alpha\mu_B B_\eta  & 0 & d_0-d_1
\end{array}\right],
\end{equation}
where we introduce the following notations 
\begin{equation}
  \begin{split}
    \alpha=(g_{\eta x}^\text{c}+g_{\eta x}^\text{v})/2,\quad \beta=(g_{\eta x}^\text{c}-g_{\eta x}^\text{v})/2, \\
\gamma=(g_{\xi y}^\text{c}-g_{\xi y}^\text{v})/2, \quad  \delta=(g_{\xi y}^\text{c}+g_{\xi y}^\text{v})/2.
  \end{split}
\end{equation}

The energies of excitonic states in both Voigt configurations calculated within such a simplified model are presented in Fig.~\ref{fig:Voi1}(a),(d), and Fig.~\ref{fig:Voi2}(a),(d) with the values of different parameters given in Appendix~C. All the observed effects are well reproduced: positions of brightened ground states of dark excitons, slight energy shift of the 1s$^{\alpha}$ state, splitting of excited states in both configurations and difference between coupling constants between bright and dark states for $ns^{\alpha}$ and $ns^{\beta}$ series. The lack of dark excited states in the $\beta$-polarized photoluminescence in $B\perp b$ configuration can be explained by the low coupling constant between $ns^{\beta}$ and D$ns^{\alpha}$ series and also the highest energy difference of all four exciton series. The fact that no brightening of the ground states of dark excitons is observed in the $\beta$-polarization can be explained by a relatively large energy difference between 1s$^{\beta}$ and D$ns^{\alpha}$/D$ns^{\beta}$ states. The proposed simplified model fits well with the experimental data presented in Fig.~\ref{fig:Voi1} and Fig.~\ref{fig:Voi2}. Nevertheless, we must admit that it still implies the use of multiple fit parameters that calls for their further confirmation.

\begin{figure*}[bt]
	\includegraphics[width=14.5cm]{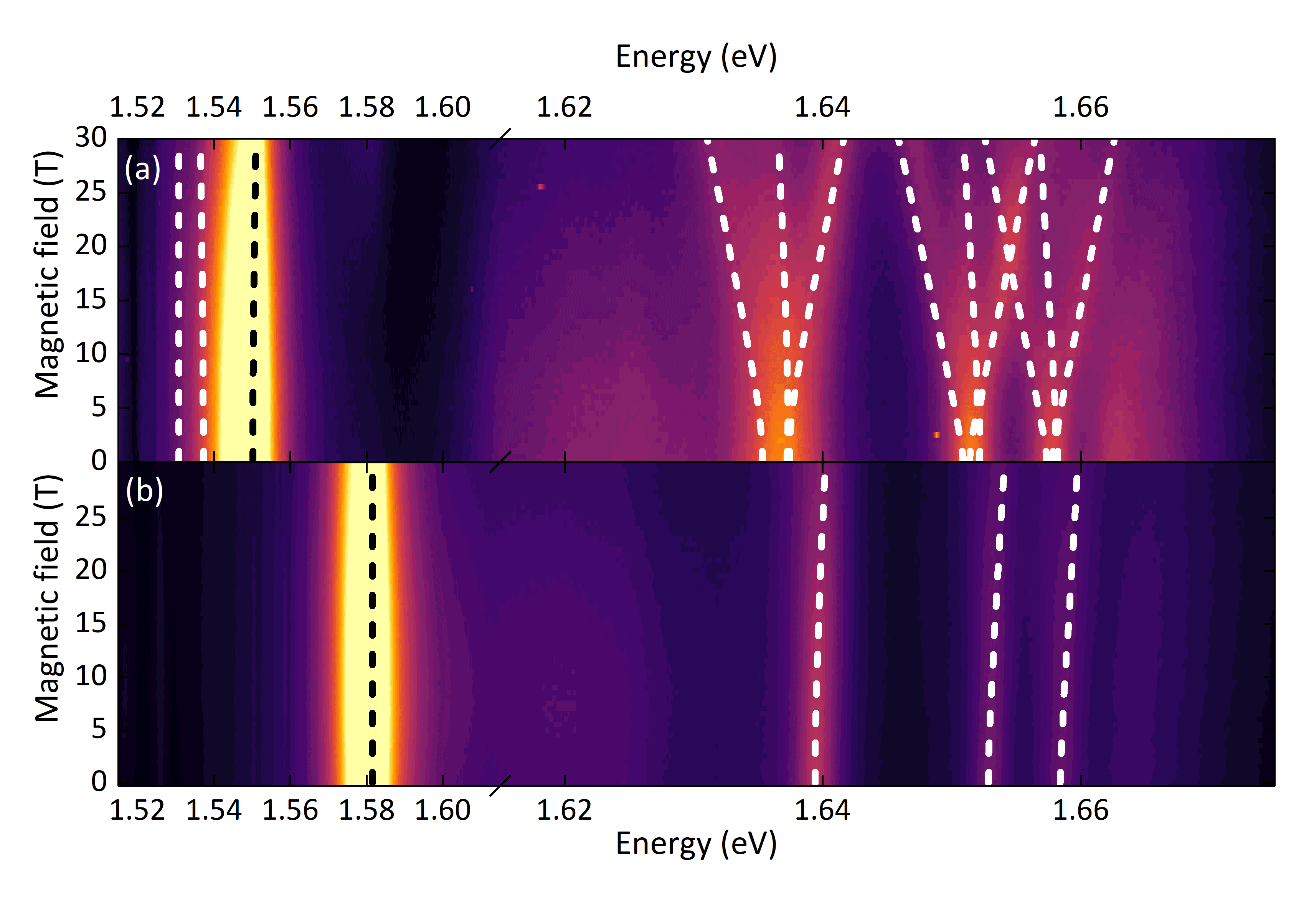}
	\caption{False-colour maps of the PL response as a function of magnetic field in the Voigt geometry with $\angle (B,b) \simeq 20^\circ$ for (a) $\alpha$- and (b) $\beta$- linear polarizations. The energy scales on both figures has been stretched in the energy region of the excited transitions. Dashed white and black lines represent transition energies obtained from the theoretical model. }
	\label{fig:Voi3}
\end{figure*}

\subsection{Magnetophotoluminescence in Voigt configuration $\angle (B,b) \simeq 20^\circ$}
\label{sec:theoryexp}

To validate the proposed model, we test it against an additional set of experimental data. In this case, the magneto-luminescence measurements in Voigt geometry were performed when the applied magnetic field $B$ is neither perpendicular nor parallel to the crystal $b$-axis, but in a defined geometry of $\angle (B,b) \simeq 20^\circ$. As shown in Fig.~\ref{fig:Voi3}., the resulting data are in fair agreement with the energy pattern of transitions expected from our model. In accordance with its predictions, when focusing on the data obtained in the configuration of $\alpha$-polarization, the apparent splitting of each excited state into three components is observed. In contrast, 1s$^{\alpha}$ and the excited states of the $ns^{\beta}$ series are only blue-shifted with the magnetic field. At the same time, the energy of the 1s$^{\beta}$ resonance does not change with the magnetic field. Nevertheless, it should be noted that the model predicts weak anticrossing between $ns^{\beta}$ series and blue-shifted dark states. However, as the magnitude of these anticrossings is significantly smaller than the linewidths of observed transitions, such effect could not be experimentally observed and was omitted here for the clarity of data presentation.

\section{Discussion \label{sec:discussion}}

The results of the magnetophotoluminescence measurements in the Voigt configurations stand as proof of the existence of two non-degenerate dark exciton series, which are coupled by the in-plane field to the bright exciton series.  Very similar effects are observed in anisotropic quantum dots of III-V compounds. This similarity can be explained by the highly anisotropic nature of bulk rhenium disulfide crystals. The layered character of this material with weak coupling between adjacent layers \cite{Tongay2014, Yu2015, Jariwala2016, Echeverry2018} leads to confinement of the carriers in two dimensions (single layer). The in-plane anisotropy, resulting from the formation of the rhenium chains, leads to even stronger confinement and a further reduction of the dimensionality of the carriers. As both valence and conduction bands are only spin-degenerate \cite{Echeverry2018}, excitons of 4 different spin configurations can be formed at a given point in the Brillouin zone. Due to the aforementioned anisotropy, the exchange interaction splits these spin states into two linear combinations of bright and two of dark states. Here, however, the fine structure splitting of the ground states of excitons is on the order of 30 meV, hence orders of magnitude larger than in the case of typical nanostructures of III-V and II-VI compounds or carbon nanotubes \cite{Seguin2005, Kowalik2007, Srivastava2008}.

It should be noted that our interpretation of excitonic resonances in ReS$_2$ is in stark contrast to the recent proposition, according to which the two linearly polarized excitons originate from two valleys: Z and K1, with both series accompanied by the dark counterparts \cite{Dhara2020AdditionalEF}. 
If that was the case, it would not be possible to achieve the in-plane-direction selective coupling of bright exciton from one valley with two dark excitons from two valleys. Our results show that $ns^{\alpha}$ and $ns^{\beta}$ bright excitons are clearly coupled by in-plane magnetic field with both D$ns^{\alpha}$ and D$ns^{\beta}$ dark excitons, and that this coupling depends on the in-plane direction of the magnetic field, implying that all these transitions have to originate from the same valley. Moreover, including two independent valleys would give a rise to at least 8 excitonic spin-states, 4 per each valley, more than observed in the experiment.
 
As presented, applying the simplified (by disregarding some of the parameters) model describing excitons in bulk rhenium disulfide allows us to explain most of the experimental results. Nevertheless, further theoretical studies are needed to firmly explain the observed polarization rules governing the optical transitions in this material. We hope that our work will stimulate more research activity on this material, allowing us to describe all the discovered phenomena fully.

\section{Summary \label{sec:summary}}

Multiple resonances can be observed in the low-temperature, zero magnetic field photoluminescence spectrum of the bulk rhenium disulfide. Here we propose their identifications as ground and excited states of two non-degenerate linearly polarized exciton series, split by the strong exchange interaction. As revealed by magnetophotoluminescence measurements performed in Voigt configuration, these bright series are accompanied by other two, non-degenerate dark exciton series. Those resonances exhibit complex interactions when magnetic field is applied, which strongly depend on the direction of the applied field. All the observed phenomena can be qualitatively described with the  application of the model including the magnetic field interaction with exciton spin states split by the exchange coupling. These results confirm that all the observed exciton species originate form the single point or two opposite points in the Brilloiun zone.
 
\section*{Acknowledgments}
The work has been supported by: EU Graphene
Flagship, CNRS via IRP “2D materials”, the ATOMOPTO project (TEAM programme of the Foundation for Polish Science, co-financed by the EU within the ERDFund), the Nano fab facility of the Institut N\'eel/CNRS/UGA, EMFL and DIR/WK/2018/07 grant from MEiN of Poland.

\section*{Appendix A. Unpolarized zero-field photoluminescence spectrum \label{0T}}

\begin{figure*}[bt]
	\includegraphics[width=14.5cm]{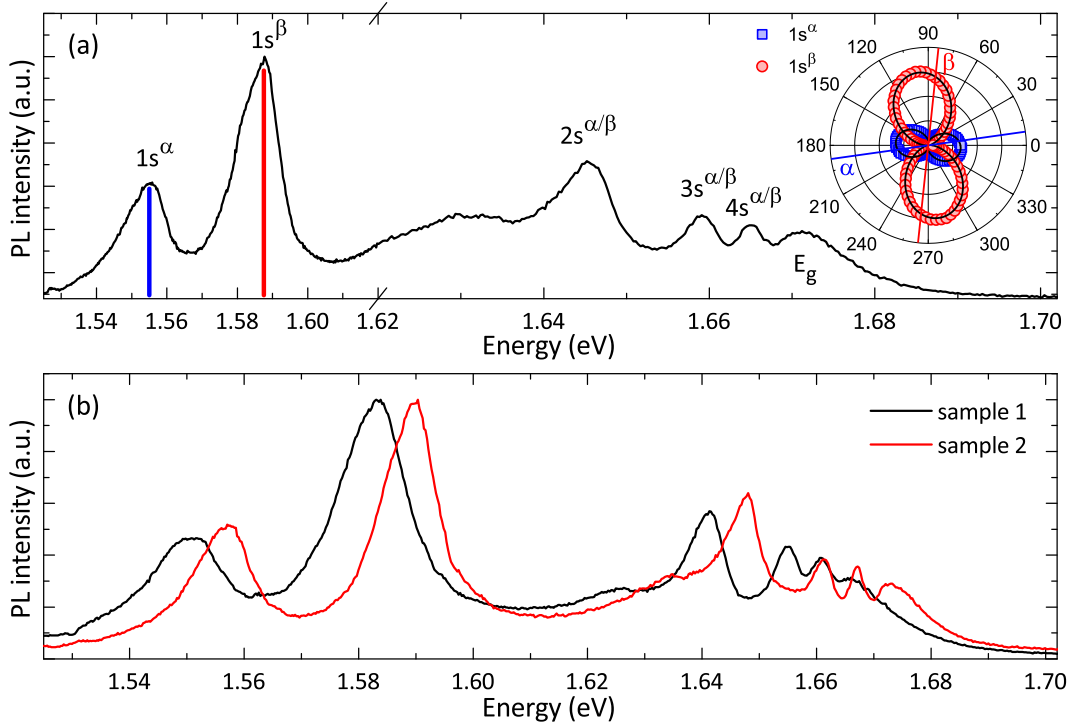}
	\caption{(a) Unpolarized low-temperature photoluminescence response of the bulk ReS$_2$ sample. The energy scale has been stretched in the energy region of the excited transitions for clarity. The inset presents the PL intensities of 1s$^{\alpha/\beta}$ transitions as a function of the angle of the linear polarization of the signal obtained at the energies marked by the blue and red vertical lines in the main figure, along with the black lines representing fits with sine function. Blue and red lines in the inset represent the $\alpha$- and $\beta$- linear polarization angles for which 1s$^{\beta}$ and 1s$^{\alpha}$ series are absent from the spectra, respectively. Blue and red line thus represent polarization angles in which only 1s$^{\alpha}$ and 1s$^{\beta}$ are visible, respectively. 
	(b) Comparison of unpolarized low-temperature photoluminescence spectra of two different bulk ReS$_2$ samples. Although the absolute energies of the observed resonances differ from sample to sample, the energy differences between subsequent emission lines remain unchanged.
	}
	\label{fig:unp0T}
\end{figure*}

The unpolarized low-temperature ($T=4.2$ K) photoluminescence (PL) response of the bulk ReS$_2$ is shown in Fig.~\ref{fig:unp0T}(a). Several emission lines can be distinguished, here denoted as: 1s$^{\alpha}$, 1s$^{\beta}$,  2s$^{\alpha/\beta}$, 3s$^{\alpha/\beta}$, 4s$^{\alpha/\beta}$ and E$_g$. As it was reported in previous works on this material, the 1s$^{\alpha}$ and 1s$^{\beta}$ transitions are almost perfectly linearly polarized\cite{Aslan2016, Urban_2018, Sim2018, Jadczak2019SR, HO2019641, Xiaofan2019}. The photoluminescence intensity at the energy indicated by the blue and red color bars, as a function of the collection angle are presented in the inset of Fig.~\ref{fig:unp0T}(a). They both follow a sine-like dependence and are rotated by 75$^\circ$ with respect to one another.  The two particular collection angles for which the intensities of 1s$^{\alpha}$ and 1s$^{\beta}$ vanish are referred to as $\beta$ and $\alpha$ polarizations, respectively.

It should be noted, that the absolute values of the observed transitions energies vary from sample to sample or even from the spot on the sample to another (as it happens also in the case of the monolayers of group VI S-TMDs), the relative energies, or the energy differences between observed resonances, are constant (see. Fig.~\ref{fig:unp0T}(b)), meaning that while the band gap varies, the exciton binding energy remains largely unchanged.

\section*{Appendix B. Magnetophotoluminescence in Faraday geometry \label{faraday}}

\begin{figure*}[bt]
	\includegraphics[width=14.5cm]{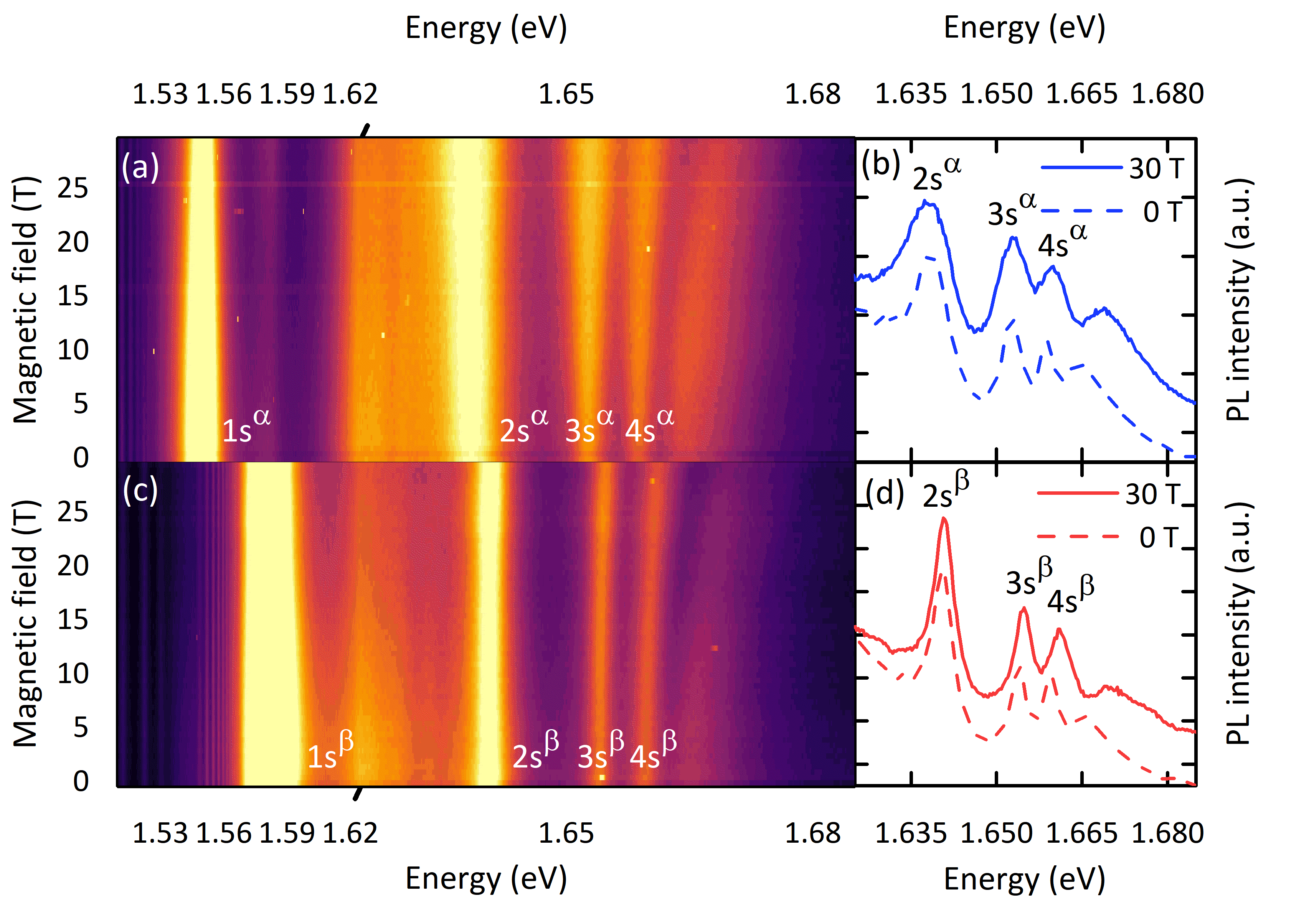}
	\caption{False-colour maps of the PL response as a function of magnetic field in the Faraday configuration for (a) $\alpha$- and (c) $\beta$- linear polarizations. The energy scales has been stretched in the energy region of the excited transitions for clarity. Comparison of the spectra obtained for (b) $\alpha$- and (d) $\beta$- linear polarizations at $B=0$~T (dashed lines) and $B=30$~T (solid lines) in the range of the excited exciton states.}
	\label{fig:Far1}	
\end{figure*}

The magnetophotoluminescence response of  bulk ReS$_2$  measured in both $\alpha$ and $\beta$ polarizations in the Faraday configuration in magnetic fields up to 30 T is shown in the figure Fig.~\ref{fig:Far1}. The magnetic field applied in such geometry only weakly affects the observed excitonic states. It has two main effects: resonances are slightly blueshifted and some of the transitions are  broadened. The observed blueshift is surprisingly weak even for the higher excited excitonic states: less than 2 meV of shift at 30~T even for 4s$^{\alpha/\beta}$ states. The brightening effect of dark excitons is not observed in this configuration.

\section*{Appendix C. Detailed theoretical description \label{apptheory}}

\subsection{Excitons from a high-symmetry point}

If the excitons arise in the vicinity of a high-symmetry point of the Brillouin zone (that is, $-\vec{k}=\vec{k}$ or $\vec{k}+\vec{b}$, with a reciprocal lattice vector~$\vec{b}$), the conduction and the valence bands must be twice degenerate (by Kramers theorem) and have definite parity which is opposite in the two bands (since there are allowed optical transitions). In the most general case of strong SOI the Kramers pair in the conduction band can be written as ($\sigma^y$~being the second Pauli matrix)
\begin{align}
&\psi_{\mathrm{c}\uparrow}(\vec{r})=e^{i\vec{k}\vec{r}}
\left(\begin{array}{c} u_{\mathrm{c}1}(\vec{r}) \\ u_{\mathrm{c}2}(\vec{r}) \end{array}\right),\\
&\psi_{\mathrm{c}\downarrow}(\vec{r})=-i\sigma^y\psi_{\mathrm{c}\uparrow}^*(\vec{r})
=e^{-i\vec{k}\vec{r}}\left(\begin{array}{c} -u_{\mathrm{c}2}^*(\vec{r}) \\ u_{\mathrm{c}1}^*(\vec{r}) \end{array}\right),
\end{align}
and similarly in the valence band with Bloch functions $u_{\mathrm{v}1}(\vec{r})$, $u_{\mathrm{v}2}(\vec{r})$. The transition dipoles of the four degenerate excitons with zero center-of-mass momentum are 
\begin{align}
&\langle\mathrm{c}\!\!\uparrow|\grad|\mathrm{v}\!\!\uparrow\rangle
=\int\left(u_{\mathrm{c}1}^*\grad{u}_{\mathrm{v}1}+u_{\mathrm{c}2}^*\grad{u}_{\mathrm{v}2}\right)d^3\vec{r},\label{upgradup=}\\
&\langle\mathrm{c}\!\!\uparrow|\grad|\mathrm{v}\!\!\downarrow\rangle
=\int\left(-u_{\mathrm{c}1}^*\grad{u}_{\mathrm{v}2}^*+u_{\mathrm{c}2}^*\grad{u}_{\mathrm{v}1}^*\right)d^3\vec{r},\label{upgraddown=}\\
&\langle\mathrm{c}\!\!\downarrow|\grad|\mathrm{v}\!\!\uparrow\rangle
=\int\left(-u_{\mathrm{c}2}\grad{u}_{\mathrm{v}1}+u_{\mathrm{c}1}\grad{u}_{\mathrm{v}2}\right)d^3\vec{r},\label{downgradup=}\\
&\langle\mathrm{c}\!\!\downarrow|\grad|\mathrm{v}\!\!\downarrow\rangle
=\int\left(u_{\mathrm{c}1}\grad{u}_{\mathrm{v}1}^*+u_{\mathrm{c}2}\grad{u}_{\mathrm{v}2}^*\right)d^3\vec{r}.
\label{downgraddown=}
\end{align}
These give two independent complex vectors or four real vectors, so one cannot make any dark linear combinations. 

If SOI is absent, the orbital and spin part separate, that is $u_{\mathrm{c}1}(\vec{r})=u_{\mathrm{c}1}^*(\vec{r})$, $u_{\mathrm{c}2}(\vec{r})=0$, and the same for the valence band. Then the second and the third line vanish immediately; they correspond to the $S^z=\pm1$ components of the triplet. The first and the last line are real and equal, so their antisymmetric linear combination is dark, corresponding to the $S^z=0$ component of the triplet.

For an $S^z$-conserving SOI, we have $u_{\mathrm{c}2}(\vec{r})=0$, $u_{\mathrm{v}2}(\vec{r})=0$, but $u_{\mathrm{c}1}(\vec{r})$ and $u_{\mathrm{v}1}(\vec{r})$ are complex. The second and the third line still vanish, but the first and the last line are complex conjugate of each other, thereby defining two independent real vectors. Then, no linear combination of the two excitons can nullify the matrix element, so each of the two excitons must remain bright in some polarization.

Let us assume that the four excitonic states are described by the same envelope function $\Psi(\vec{r}_e,\vec{r}_h)$, and are labelled by the spin projections, $|s_\mathrm{c},s_\mathrm{v}\rangle$ with $s_\mathrm{c},s_\mathrm{v}=\uparrow,\downarrow$ (note that the hole spin is $-s_\mathrm{v}$). The exchange matrix elements of the Coulomb interaction $V(\vec{r}-\vec{r}')$ between two excitonic states are given by
\begin{align}
V_{s_\mathrm{c},s_\mathrm{v},s_\mathrm{c}',s_\mathrm{v}'}=&{}-\int{d}^3\vec{r}\,{d}^3\vec{r}'\,
\left[\Psi^*(\vec{r},\vec{r})\,\psi^\dagger_{\mathrm{v}s_\mathrm{v}}(\vec{r})\,\psi_{\mathrm{c}s_\mathrm{c}}(\vec{r})\right]\times{}\nonumber\\
{}&{}\times
V(\vec{r}-\vec{r}')\left[\Psi(\vec{r}',\vec{r}')\,\psi^\dagger_{\mathrm{c}s_\mathrm{c}'}(\vec{r}')\,\psi_{\mathrm{v}s_\mathrm{v}'}(\vec{r}')\right],
\end{align}
and define a $4\times4$ matrix. Which of the matrix elements are non-zero, can be understood quite analogously to the transition dipole moments. Indeed, all conclusions regarding Eqs.~(\ref{upgradup=}),~(\ref{upgraddown=}),~(\ref{downgradup=}),~(\ref{downgraddown=}) remain valid if one replaces $\grad$ by an arbitrary real function of $\vec{r}$.
So for vanishing SOI the exchange matrix is diagonal in the singlet-triplet basis, with only one non-zero matrix element (singlet-singlet).
For a generic SOI, all matrix elements are generally speaking non-zero.
For the case of an $S^z$-conserving SOI, we have a $2\times2$ matrix in the space of the two bright excitons $|\uparrow\uparrow\rangle$, $|\downarrow\downarrow\rangle$:
\begin{align}
V_\mathrm{b}={}&{}-\int{d}^3\vec{r}\,{d}^3\vec{r}'\,V(\vec{r}-\vec{r}')
\left(\begin{array}{cc}
\rho^*(\vec{r})\,\rho(\vec{r}') &
\rho^*(\vec{r})\,\rho^*(\vec{r}') \\
\rho(\vec{r})\,\rho(\vec{r}') &
\rho(\vec{r})\,\rho^*(\vec{r}') \\
\end{array}\right)\nonumber\\
\equiv{}&{}\left(\begin{array}{cc} b_0 & b_1 \\ b_1^* & b_0 \end{array}\right),
\end{align}
where we denoted $\rho(\vec{r})=u_{\mathrm{c}1}^*(\vec{r})\,u_{\mathrm{v}1}(\vec{r})$.
If we write $b_1=|b_1|e^{2i\phi_b}$, then the eigenvectors are 
\begin{equation}
|\pm\rangle=\frac{e^{i\phi_b}|\!\uparrow\uparrow\rangle\pm{e}^{-i\phi_b}|\!\downarrow\downarrow\rangle}{\sqrt2}
\end{equation}
with the eigenvalues $b_0\pm|b_1|$. The matrix elements of dipolar transition into these excitonic eigenstates are purely real or purely imaginary:
\begin{equation}
\vec{D}_\pm=
\frac{e^{-i\phi_b}}{\sqrt{2}}\int{d}^3\vec{r}\,u_{\mathrm{c}1}^*\grad{u}_{\mathrm{v}1}\pm
\frac{e^{i\phi_b}}{\sqrt{2}}\int{d}^3\vec{r}\,u_{\mathrm{c}1}\grad{u}^*_{\mathrm{v}1},
\end{equation}
and thus determine two linear polarizations, in agreement with the experimental observations.

The wave functions $u_\mathrm{c1}(\vec{r})$ and $u_\mathrm{v1}(\vec{r})$ are essentially complex, i.~e., they cannot be made real with a simple global phase factor $e^{i\phi_\mathrm{c}}$,  $e^{i\phi_\mathrm{v}}$. Still we have a freedom in choosing these phase factors, and we can use this freedom to make the matrix elements $b_1,d_1$ in Eq.~(\ref{eq:exchange_Hamiltonian}) real and positive. Indeed, if for some choice of $u_\mathrm{c1}(\vec{r})$ and $u_\mathrm{v1}(\vec{r})$ we have $b_1=|b_1|e^{2i\phi_\mathrm{b}}$, $d_1=|d_1|e^{2i\phi_\mathrm{d}}$, we can choose new wave functions $\tilde{u}_\mathrm{c1}(\vec{r})=e^{-i(\phi_\mathrm{b}+\phi_\mathrm{d})/2}{u}_\mathrm{c1}(\vec{r})$ and $\tilde{u}_\mathrm{v1}(\vec{r})=e^{i(\phi_\mathrm{b}-\phi_\mathrm{d})/2}{u}_\mathrm{v1}(\vec{r})$ which will give new matrix elements $\tilde{b}_1=|b_1|$,  $\tilde{d}_1=|d_1|$. Note that fixing the phases of $u_\mathrm{c1}(\vec{r})$ and $u_\mathrm{v1}(\vec{r})$ implies fixing the $x$ and $y$ directions for the electron spin; indeed, a spin rotation around the $z$ axis, $e^{i\sigma^z\phi/2}$ would affect the wave function phases, which is no longer allowed. 

\subsection{Excitons from two opposite valleys}

Let us now consider two valleys, located in some opposite points $\pm\vec{k}$ of the Brillouin zone. In the most general case of strong SOI we can write one of the two degenerate states at the $\vec{k}$ point (which we arbitrarily label by $\uparrow$) in the conduction band as
\begin{equation}
\psi_{\mathrm{c}\uparrow\vec{k}}(\vec{r})=e^{i\vec{k}\vec{r}}
\left(\begin{array}{c} u_{\mathrm{c}1}(\vec{r}) \\ u_{\mathrm{c}1}(\vec{r}) \end{array}\right).
\end{equation}
Then the $\uparrow$ state at $-\vec{k}$ can be obtained from it by space inversion,
\begin{equation}
\psi_{\mathrm{c}\uparrow-\vec{k}}(\vec{r})=\psi_{\mathrm{c}\uparrow\vec{k}}(-\vec{r})
=e^{-i\vec{k}\vec{r}}
\left(\begin{array}{c} u_{\text{c}_1}(-\mathbf{r}) \\ u_{\text{c}_2}(-\mathbf{r}) \end{array}\right),
\end{equation}
while applying the time reversal to $\psi_{\mathrm{c}\uparrow\vec{k}}(\vec{r})$ we obtain the $\downarrow$ state at $-\vec{k}$:
\begin{equation}
\psi_{\mathrm{c}\downarrow-\vec{k}}(\vec{r})=
-i\sigma^y\psi_{\mathrm{c}\uparrow\vec{k}}^*(\vec{r})=
e^{-i\vec{k}\vec{r}}
\left(\begin{array}{c} -u_{\text{c}_2}^*(\mathbf{r})) \\ u_{\text{c}_1}^*(\mathbf{r}) \end{array}\right),
\end{equation}
Applying space inversion to the latter state, we obtain the last one, $\downarrow$ at $\vec{k}$:
\begin{equation}
\psi_{\mathrm{c}\downarrow\vec{k}}(\vec{r})=
\psi_{\mathrm{c}\downarrow-\vec{k}}(-\vec{r})=
e^{i\vec{k}\vec{r}}\left(\begin{array}{c} -u_{\text{c}_2}^*(-\mathbf{r}) \\ u_{\text{c}_1}^*(-\mathbf{r}) \end{array}\right).
\end{equation}
By the same construction one can build four degenerate states in the valence band.
 
Out of these single-electron states, one can construct 16 different excitonic species.
Half of them correspond to electrons and holes from opposite valleys; their optical transitions are forbidden by momentum conservation and can be removed from the consideration. The remaining 8 exciton species are formed by electrons and holes from the same valley. Similarly to the one-valley case, these species can be labelled as  $|s_\mathrm{c},s_\mathrm{v},\upsilon\rangle$, where $s_\mathrm{c},s_\mathrm{v}=\uparrow,\downarrow$ labels the spin projections in the conduction and the valence band, while $\upsilon=\pm$ labels the valleys $\pm\vec{k}$.     

These states, however, do not have a definite parity, required by the $C_i$ symmetry of the crystal; one has to form even and odd linear combinations: $|s_\mathrm{c}s_\mathrm{v},\mathrm{e}\rangle=(|s_\mathrm{c}s_\mathrm{v},+\rangle+
|s_\mathrm{c}s_\mathrm{v},-\rangle)/\sqrt{2}$, 
$|s_\mathrm{c}s_\mathrm{v},\mathrm{o}\rangle=(|s_\mathrm{c}s_\mathrm{v},+\rangle-
|s_\mathrm{c}s_\mathrm{v},-\rangle)/\sqrt{2}$, respectively. 
For the even states, the dipole transition is forbidden by parity (since the crystal's ground state is even). It is now straightforward to check directly that for the four odd-parity excitons we arrive at the same conclusions regarding the exchange coupling and polarization, as for the single-valley case.

\subsection{Eigenvalues}

The analytic expressions for the eigenvalues of the simplified Hamiltonian Eq.~(\ref{eq:simpl_Hamiltonian}) can be found for the cases of the magnetic field applied parallel  or perpendicular  to the $b$-axis of the crystal.

In the first case ($B\parallel b$) the spectrum is:
\begin{widetext}
\begin{align}\label{eq:eigen1}
\begin{split}
E_\text{b}^+(B_\eta)=&\frac{b_0+b_1+d_0+d_1}{2}+\sqrt{\Big(\frac{b_0+b_1-d_0-d_1}{2}\Big)^2+\beta^2\mu_B^2B_\eta^2}, \\
E_\text{d}^+(B_\eta)=&\frac{b_0+b_1+d_0+d_1}{2}-\sqrt{\Big(\frac{b_0+b_1-d_0-d_1}{2}\Big)^2+\beta^2\mu_B^2B_\eta^2}, \\
E_\text{b}^-(B_\eta)=&\frac{b_0-b_1+d_0-d_1}{2}+\sqrt{\Big(\frac{b_0-b_1-d_0+d_1}{2}\Big)^2+\alpha^2\mu_B^2B_\eta^2}, \\
E_\text{d}^-(B_\eta)=&\frac{b_0-b_1+d_0-d_1}{2}-\sqrt{\Big(\frac{b_0-b_1-d_0+d_1}{2}\Big)^2+\alpha^2\mu_B^2B_\eta^2}.
\end{split}
\end{align} 
\end{widetext}
For large magnetic field $B_\eta$ behavior of the excitonic spectrum  
allows to evaluate the parameters of the Hamiltonian 
\begin{align}\label{eq:limit1}
\begin{split}
|\beta|=\frac{E_\text{b}^+(B_\eta)-E_\text{d}^+(B_\eta)}{2\mu_B B_\eta},\\
|\alpha|=\frac{E_\text{d}^-(B_\eta)-E_\text{d}^-(B_\eta)}{2\mu_B B_\eta}.
\end{split}
\end{align}

For the second case ($B\perp b$) we have:
\begin{widetext}
\begin{align}\label{eq:eigen2}
\begin{split}
E_\text{b}^+(B_\xi)=&\frac{b_0+b_1+d_0-d_1}{2}+
\sqrt{\Big(\frac{b_0+b_1-d_0+d_1}{2}\Big)^2+\delta^2\mu_B^2B_\xi^2}, \\
E_\text{d}^-(B_\xi)=&\frac{b_0+b_1+d_0-d_1}{2}-
\sqrt{\Big(\frac{b_0+b_1-d_0+d_1}{2}\Big)^2+\delta^2\mu_B^2B_\xi^2}, \\
E_\text{b}^-(B_\xi)=&\frac{b_0-b_1+d_0+d_1}{2}+
\sqrt{\Big(\frac{b_0-b_1-d_0-d_1}{2}\Big)^2+\gamma^2\mu_B^2B_\xi^2}, \\  
E_\text{d}^+(B_\xi)=&\frac{b_0-b_1+d_0+d_1}{2}-
\sqrt{\Big(\frac{b_0-b_1-d_0-d_1}{2}\Big)^2+\gamma^2\mu_B^2B_\xi^2}.
\end{split}
\end{align}
\end{widetext}
The absolute value of the parameter $\delta$ can be found in the limit of the large magnetic field $B_\xi$
\begin{align}\label{eq:limit2}
\begin{split}
|\delta|=\frac{E_\text{b}^+(B_\xi)-E_\text{d}^-(B_\xi)}{2\mu_BB_\xi},\\
|\gamma|=\frac{E_\text{b}^+(B_\xi)+E_\text{d}^-(B_\xi)}{2\mu_BB_\xi}.
\end{split}
\end{align}

\begingroup
\def\arraystretch{2}
\begin{table*}[]
\caption{Values of the parameters estimated from the experiment and used for the the calculations.}
\begin{tabular*}{0.75\textwidth}{@{\extracolsep{\fill}}crrrrrrr@{}}
\hline\hline
$n$ & \multicolumn{1}{c}{$b_0-d_0$ (meV)} & \multicolumn{1}{c}{$b_1$ (meV)} & \multicolumn{1}{c}{$d_1$ (meV)} & \multicolumn{1}{c}{$g_{\eta x}^\text{c}$} & \multicolumn{1}{c}{$g_{\eta x}^\text{v}$} & \multicolumn{1}{c}{$g_{\xi y}^\text{c}$} & \multicolumn{1}{c}{$g_{\xi y}^\text{v}$} \\
 \hline  
1   & 31.80                          & 16.12 & 3.18 & 2.8                                   & 1.6               & -2.4                               & 1.3           \\
2   & 1.94                           & 1.08 & 1.04 & 4.0                                   & 2.2               & -2.2                               & 1.2           \\
3   & 0.60                           & 0.72                           & 0.66                           & 4.1                                   & 2.2               & -2.2                               & 1.2           \\
4   & 0.31                           & 0.30 & 0.43 & 3.9                                   & 2.1               & -1.9                               & 1.0 \\   \hline\hline      
\end{tabular*}
	\label{tab:par}
\end{table*}

\section*{Appendix D. Parameters} \label{theory:parameters}

\begin{figure}[bt]
	\includegraphics[width=\columnwidth]{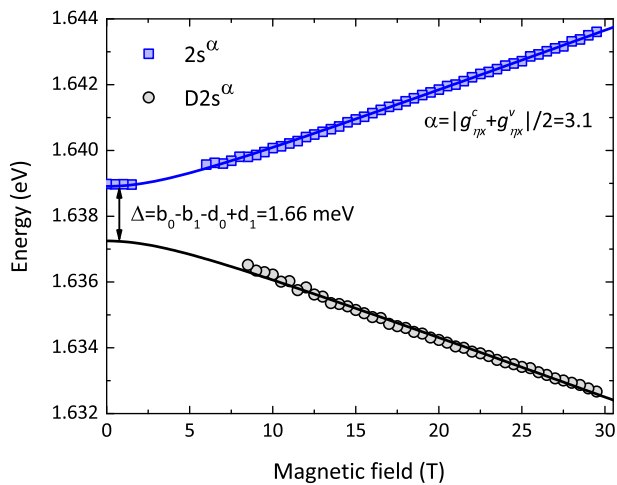}
	\caption{Energies of the $\unit{2s}^{\alpha}$ and $\unit{D2s}^{\alpha}$ as a function of the in-plane magnetic field $B\parallel b$ along with the best fit to the corresponding eigenvalues Eq.~(\ref{eq:eigen1}) of the model Eq.~(\ref{eq:simpl_Hamiltonian}).}
	\label{fig:parest}
\end{figure}

To be able to simulate the experimental results with the model described above, one needs to find values of $g$-factors and exchange Hamiltonian parameters. To do so, the energies of the bright and dark states as a function of magnetic field were fitted using the eigenvalues Eqs.~(\ref{eq:eigen1}),~(\ref{eq:eigen2}) of simplified Hamiltonian Eq.~(\ref{eq:simpl_Hamiltonian}). The example of such fitting was presented in the Fig.~\ref{fig:parest}. This procedure was applied to measurements in the $\alpha$-polarization in both $B\parallel b$ and $B\perp b$ configurations. Combination of the results from these two measurements allowed to estimate values of the exchange splitting parameters. However, similar approach could not be implemented for the measurements in the $\beta$-polarization, as the effect of magnetic field was too weak. In that case high field limits were used: Eqs.~(\ref{eq:limit1}),~(\ref{eq:limit2}). These very rough estimations allow us to obtain all the necessary parameters, summarized in the Table~\ref{tab:par}.

It should be noted, that the extracted values of the exchange parameters in the case of excited states are significantly smaller than the typical linewidths of the peaks ($\approx 5$ meV). Thus, these values should not be taken as exact. 

\section*{Appendix E. Methods}

\begin{figure*}[bt]
	\includegraphics[width=9cm]{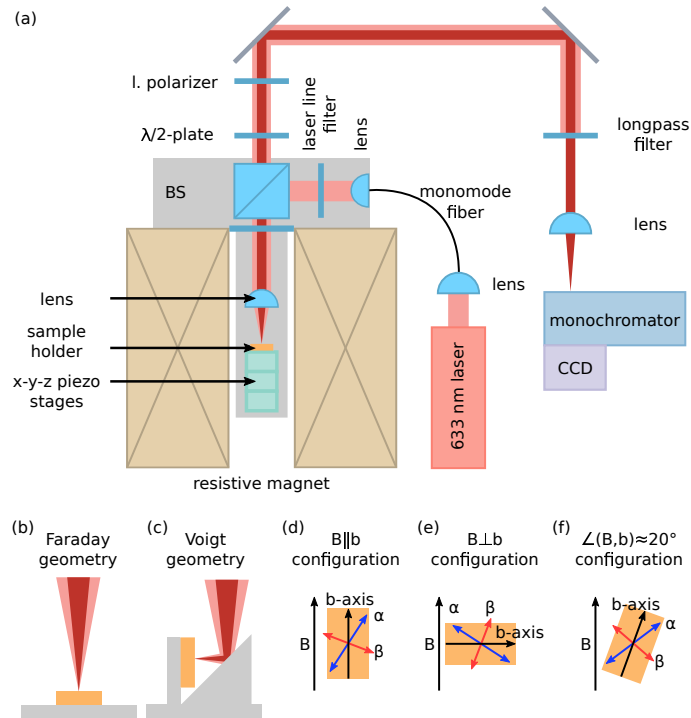}
	\caption{(a) Sketch of the optical setup for magnetophotoluminescence measurements. The sample placement in (b) Faraday and (c) Voigt geometries. The diagrams presenting the magnetic field vector B, crystallographic b-axis, and $\alpha$- and $\beta$-polarization axes for the measurements in (d) $B\parallel b$, (e) $B\perp b$ and (f) $\angle (B,b) \simeq 20^\circ$ configurations in Voigt geometry. The angles of $\alpha$- and $\beta$-polarization axes were determined according to \cite{Jadczak2019SR}.}
	\label{fig:setup}
\end{figure*}

Low-temperature ($T$=4.6 K) zero-field experiments were conducted using continuous wave laser ($\lambda$=515~nm) as excitation. The sample was placed on a cold finger of a continuous flow cryostat. The excitation light ($P_{exc}=1$~mW) was focused by means
of a 50x long-working distance objective with a 0.5 numerical aperture producing a spot of about 1 µm. The
photoluminescence signal was collected via the same objective and went through $\lambda$/2-plate, which was used to change linear polarization, and linear polarization analyser, which was kept in the same position. The measured angle of polarization is arbitrary and does not reflect the angle between light polarization and any of the crystal axis. The collected signal was dispersed through a 0.5 m long monochromator and detected by a CCD camera.

The magnetophotoluminescence measurements were performed using free-beam-optics insert probe kept in liquid helium cryostat placed inside 30 T resistive coil (see Fig.~\ref{fig:setup}(a)). The excitation light ($\lambda$=633~nm) was focused by means of an aspheric lens on the sample placed on top of a x-y-z piezo-stage. The excitation power on the sample was estimated to be equal to $P_{exc}=1$~mW. The signal was collected by the same lens and went through the $\lambda$/2-plate and linear polarizer. The collected signal was dispersed through a 0.5 m long monochromator and detected by a CCD camera.

The alignment of the samples versus the direction of magnetic field was achieved by using the linear breaks of cracks that can be seen on the surface of the ReS$_2$ crystals. These cracks indicate the rhenium chains in the material, as the strong Re-Re bonds lead to the preferential breaking along them \cite{Chenet2015,LIANG200994}. Since the rhenium chains are parallel to the $b$-axis of the crystal, the linear cracks allow to determine the crystallographic orientation of the sample. The correctness of the alignment was then checked using a polarizer and a half-wave plate.

\bibliography{bibliography}

\begin{thebibliography}{50}%
\makeatletter
\providecommand \@ifxundefined [1]{%
 \@ifx{#1\undefined}
}%
\providecommand \@ifnum [1]{%
 \ifnum #1\expandafter \@firstoftwo
 \else \expandafter \@secondoftwo
 \fi
}%
\providecommand \@ifx [1]{%
 \ifx #1\expandafter \@firstoftwo
 \else \expandafter \@secondoftwo
 \fi
}%
\providecommand \natexlab [1]{#1}%
\providecommand \enquote  [1]{``#1''}%
\providecommand \bibnamefont  [1]{#1}%
\providecommand \bibfnamefont [1]{#1}%
\providecommand \citenamefont [1]{#1}%
\providecommand \href@noop [0]{\@secondoftwo}%
\providecommand \href [0]{\begingroup \@sanitize@url \@href}%
\providecommand \@href[1]{\@@startlink{#1}\@@href}%
\providecommand \@@href[1]{\endgroup#1\@@endlink}%
\providecommand \@sanitize@url [0]{\catcode `\\12\catcode `\$12\catcode
  `\&12\catcode `\#12\catcode `\^12\catcode `\_12\catcode `\%12\relax}%
\providecommand \@@startlink[1]{}%
\providecommand \@@endlink[0]{}%
\providecommand \url  [0]{\begingroup\@sanitize@url \@url }%
\providecommand \@url [1]{\endgroup\@href {#1}{\urlprefix }}%
\providecommand \urlprefix  [0]{URL }%
\providecommand \Eprint [0]{\href }%
\providecommand \doibase [0]{https://doi.org/}%
\providecommand \selectlanguage [0]{\@gobble}%
\providecommand \bibinfo  [0]{\@secondoftwo}%
\providecommand \bibfield  [0]{\@secondoftwo}%
\providecommand \translation [1]{[#1]}%
\providecommand \BibitemOpen [0]{}%
\providecommand \bibitemStop [0]{}%
\providecommand \bibitemNoStop [0]{.\EOS\space}%
\providecommand \EOS [0]{\spacefactor3000\relax}%
\providecommand \BibitemShut  [1]{\csname bibitem#1\endcsname}%
\let\auto@bib@innerbib\@empty
\bibitem [{\citenamefont {Chhowalla}\ \emph {et~al.}(2013)\citenamefont
  {Chhowalla}, \citenamefont {Shin}, \citenamefont {Eda}, \citenamefont {Li},
  \citenamefont {Loh},\ and\ \citenamefont {Zhang}}]{Chhowalla2013}%
  \BibitemOpen
  \bibfield  {author} {\bibinfo {author} {\bibfnamefont {M.}~\bibnamefont
  {Chhowalla}}, \bibinfo {author} {\bibfnamefont {H.~S.}\ \bibnamefont {Shin}},
  \bibinfo {author} {\bibfnamefont {G.}~\bibnamefont {Eda}}, \bibinfo {author}
  {\bibfnamefont {L.-J.}\ \bibnamefont {Li}}, \bibinfo {author} {\bibfnamefont
  {K.~P.}\ \bibnamefont {Loh}},\ and\ \bibinfo {author} {\bibfnamefont
  {H.}~\bibnamefont {Zhang}},\ }\href {https://doi.org/10.1038/nchem.1589}
  {\bibfield  {journal} {\bibinfo  {journal} {Nature Chemistry}\ }\textbf
  {\bibinfo {volume} {5}},\ \bibinfo {pages} {263} (\bibinfo {year}
  {2013})}\BibitemShut {NoStop}%
\bibitem [{\citenamefont {Kertesz}\ and\ \citenamefont
  {Hoffmann}(1984)}]{Kertesz1984}%
  \BibitemOpen
  \bibfield  {author} {\bibinfo {author} {\bibfnamefont {M.}~\bibnamefont
  {Kertesz}}\ and\ \bibinfo {author} {\bibfnamefont {R.}~\bibnamefont
  {Hoffmann}},\ }\href {https://doi.org/10.1021/ja00324a012} {\bibfield
  {journal} {\bibinfo  {journal} {Journal of the American Chemical Society}\
  }\textbf {\bibinfo {volume} {106}},\ \bibinfo {pages} {3453} (\bibinfo {year}
  {1984})}\BibitemShut {NoStop}%
\bibitem [{\citenamefont {Lamfers}\ \emph {et~al.}(1996)\citenamefont
  {Lamfers}, \citenamefont {Meetsma}, \citenamefont {Wiegers},\ and\
  \citenamefont {de~Boer}}]{LAMFERS199634}%
  \BibitemOpen
  \bibfield  {author} {\bibinfo {author} {\bibfnamefont {H.-J.}\ \bibnamefont
  {Lamfers}}, \bibinfo {author} {\bibfnamefont {A.}~\bibnamefont {Meetsma}},
  \bibinfo {author} {\bibfnamefont {G.}~\bibnamefont {Wiegers}},\ and\ \bibinfo
  {author} {\bibfnamefont {J.}~\bibnamefont {de~Boer}},\ }\href
  {https://doi.org/https://doi.org/10.1016/0925-8388(96)02313-4} {\bibfield
  {journal} {\bibinfo  {journal} {Journal of Alloys and Compounds}\ }\textbf
  {\bibinfo {volume} {241}},\ \bibinfo {pages} {34 } (\bibinfo {year}
  {1996})}\BibitemShut {NoStop}%
\bibitem [{\citenamefont {Murray}\ \emph {et~al.}(1994)\citenamefont {Murray},
  \citenamefont {Kelty}, \citenamefont {Chianelli},\ and\ \citenamefont
  {Day}}]{Murray1994}%
  \BibitemOpen
  \bibfield  {author} {\bibinfo {author} {\bibfnamefont {H.~H.}\ \bibnamefont
  {Murray}}, \bibinfo {author} {\bibfnamefont {S.~P.}\ \bibnamefont {Kelty}},
  \bibinfo {author} {\bibfnamefont {R.~R.}\ \bibnamefont {Chianelli}},\ and\
  \bibinfo {author} {\bibfnamefont {C.~S.}\ \bibnamefont {Day}},\ }\href
  {https://doi.org/10.1021/ic00097a037} {\bibfield  {journal} {\bibinfo
  {journal} {Inorganic Chemistry}\ }\textbf {\bibinfo {volume} {33}},\ \bibinfo
  {pages} {4418} (\bibinfo {year} {1994})}\BibitemShut {NoStop}%
\bibitem [{\citenamefont {Ho}\ \emph {et~al.}(1998{\natexlab{a}})\citenamefont
  {Ho}, \citenamefont {Huang}, \citenamefont {Tiong},\ and\ \citenamefont
  {Liao}}]{PhysRevB.58.16130}%
  \BibitemOpen
  \bibfield  {author} {\bibinfo {author} {\bibfnamefont {C.~H.}\ \bibnamefont
  {Ho}}, \bibinfo {author} {\bibfnamefont {Y.~S.}\ \bibnamefont {Huang}},
  \bibinfo {author} {\bibfnamefont {K.~K.}\ \bibnamefont {Tiong}},\ and\
  \bibinfo {author} {\bibfnamefont {P.~C.}\ \bibnamefont {Liao}},\ }\href
  {https://doi.org/10.1103/PhysRevB.58.16130} {\bibfield  {journal} {\bibinfo
  {journal} {Phys. Rev. B}\ }\textbf {\bibinfo {volume} {58}},\ \bibinfo
  {pages} {16130} (\bibinfo {year} {1998}{\natexlab{a}})}\BibitemShut {NoStop}%
\bibitem [{\citenamefont {Lin}\ \emph {et~al.}(2011)\citenamefont {Lin},
  \citenamefont {Huang}, \citenamefont {Wu}, \citenamefont {Lin}, \citenamefont
  {Huang},\ and\ \citenamefont {Tiong}}]{lin2011}%
  \BibitemOpen
  \bibfield  {author} {\bibinfo {author} {\bibfnamefont {D.~Y.}\ \bibnamefont
  {Lin}}, \bibinfo {author} {\bibfnamefont {T.~P.}\ \bibnamefont {Huang}},
  \bibinfo {author} {\bibfnamefont {F.~L.}\ \bibnamefont {Wu}}, \bibinfo
  {author} {\bibfnamefont {C.~M.}\ \bibnamefont {Lin}}, \bibinfo {author}
  {\bibfnamefont {Y.~S.}\ \bibnamefont {Huang}},\ and\ \bibinfo {author}
  {\bibfnamefont {K.~K.}\ \bibnamefont {Tiong}},\ }in\ \href
  {https://doi.org/10.4028/www.scientific.net/SSP.170.135} {\emph {\bibinfo
  {booktitle} {Solid Compounds of Transition Elements I}}},\ \bibinfo {series}
  {Solid State Phenomena}, Vol.\ \bibinfo {volume} {170}\ (\bibinfo
  {publisher} {Trans Tech Publications Ltd},\ \bibinfo {year} {2011})\ pp.\
  \bibinfo {pages} {135--138}\BibitemShut {NoStop}%
\bibitem [{\citenamefont {Ho}\ \emph {et~al.}(2007)\citenamefont {Ho},
  \citenamefont {Hsieh}, \citenamefont {Wu}, \citenamefont {Huang},\ and\
  \citenamefont {Tiong}}]{HO2007245}%
  \BibitemOpen
  \bibfield  {author} {\bibinfo {author} {\bibfnamefont {C.}~\bibnamefont
  {Ho}}, \bibinfo {author} {\bibfnamefont {M.}~\bibnamefont {Hsieh}}, \bibinfo
  {author} {\bibfnamefont {C.}~\bibnamefont {Wu}}, \bibinfo {author}
  {\bibfnamefont {Y.}~\bibnamefont {Huang}},\ and\ \bibinfo {author}
  {\bibfnamefont {K.}~\bibnamefont {Tiong}},\ }\href
  {https://doi.org/https://doi.org/10.1016/j.jallcom.2006.08.359} {\bibfield
  {journal} {\bibinfo  {journal} {Journal of Alloys and Compounds}\ }\textbf
  {\bibinfo {volume} {442}},\ \bibinfo {pages} {245 } (\bibinfo {year}
  {2007})},\ \bibinfo {note} {{Proceedings} of the 15th International
  Conference on Solid Compounds of Transition Elements}\BibitemShut {NoStop}%
\bibitem [{\citenamefont {Splendiani}\ \emph {et~al.}(2010)\citenamefont
  {Splendiani}, \citenamefont {Sun}, \citenamefont {Zhang}, \citenamefont {Li},
  \citenamefont {Kim}, \citenamefont {Chim}, \citenamefont {Galli},\ and\
  \citenamefont {Wang}}]{splendiani2010}%
  \BibitemOpen
  \bibfield  {author} {\bibinfo {author} {\bibfnamefont {A.}~\bibnamefont
  {Splendiani}}, \bibinfo {author} {\bibfnamefont {L.}~\bibnamefont {Sun}},
  \bibinfo {author} {\bibfnamefont {Y.}~\bibnamefont {Zhang}}, \bibinfo
  {author} {\bibfnamefont {T.}~\bibnamefont {Li}}, \bibinfo {author}
  {\bibfnamefont {J.}~\bibnamefont {Kim}}, \bibinfo {author} {\bibfnamefont
  {C.-Y.}\ \bibnamefont {Chim}}, \bibinfo {author} {\bibfnamefont
  {G.}~\bibnamefont {Galli}},\ and\ \bibinfo {author} {\bibfnamefont
  {F.}~\bibnamefont {Wang}},\ }\href {https://doi.org/10.1021/nl903868w}
  {\bibfield  {journal} {\bibinfo  {journal} {Nano Letters}\ }\textbf {\bibinfo
  {volume} {10}},\ \bibinfo {pages} {1271} (\bibinfo {year}
  {2010})}\BibitemShut {NoStop}%
\bibitem [{\citenamefont {Mak}\ \emph {et~al.}(2010)\citenamefont {Mak},
  \citenamefont {Lee}, \citenamefont {Hone}, \citenamefont {Shan},\ and\
  \citenamefont {Heinz}}]{mak2010}%
  \BibitemOpen
  \bibfield  {author} {\bibinfo {author} {\bibfnamefont {K.~F.}\ \bibnamefont
  {Mak}}, \bibinfo {author} {\bibfnamefont {C.}~\bibnamefont {Lee}}, \bibinfo
  {author} {\bibfnamefont {J.}~\bibnamefont {Hone}}, \bibinfo {author}
  {\bibfnamefont {J.}~\bibnamefont {Shan}},\ and\ \bibinfo {author}
  {\bibfnamefont {T.~F.}\ \bibnamefont {Heinz}},\ }\href
  {https://doi.org/10.1103/PhysRevLett.105.136805} {\bibfield  {journal}
  {\bibinfo  {journal} {Phys. Rev. Lett.}\ }\textbf {\bibinfo {volume} {105}},\
  \bibinfo {pages} {136805} (\bibinfo {year} {2010})}\BibitemShut {NoStop}%
\bibitem [{\citenamefont {Ellis}\ \emph {et~al.}(2011)\citenamefont {Ellis},
  \citenamefont {Lucero},\ and\ \citenamefont {Scuseria}}]{Ellis2011}%
  \BibitemOpen
  \bibfield  {author} {\bibinfo {author} {\bibfnamefont {J.~K.}\ \bibnamefont
  {Ellis}}, \bibinfo {author} {\bibfnamefont {M.~J.}\ \bibnamefont {Lucero}},\
  and\ \bibinfo {author} {\bibfnamefont {G.~E.}\ \bibnamefont {Scuseria}},\
  }\href {https://doi.org/10.1063/1.3672219} {\bibfield  {journal} {\bibinfo
  {journal} {Applied Physics Letters}\ }\textbf {\bibinfo {volume} {99}},\
  \bibinfo {pages} {261908} (\bibinfo {year} {2011})}\BibitemShut {NoStop}%
\bibitem [{\citenamefont {Tongay}\ \emph {et~al.}(2014)\citenamefont {Tongay},
  \citenamefont {Sahin}, \citenamefont {Ko}, \citenamefont {Luce},
  \citenamefont {Fan}, \citenamefont {Liu}, \citenamefont {Zhou}, \citenamefont
  {Huang}, \citenamefont {Ho}, \citenamefont {Yan}, \citenamefont {Ogletree},
  \citenamefont {Aloni}, \citenamefont {Ji}, \citenamefont {Li}, \citenamefont
  {Li}, \citenamefont {Peeters},\ and\ \citenamefont {Wu}}]{Tongay2014}%
  \BibitemOpen
  \bibfield  {author} {\bibinfo {author} {\bibfnamefont {S.}~\bibnamefont
  {Tongay}}, \bibinfo {author} {\bibfnamefont {H.}~\bibnamefont {Sahin}},
  \bibinfo {author} {\bibfnamefont {C.}~\bibnamefont {Ko}}, \bibinfo {author}
  {\bibfnamefont {A.}~\bibnamefont {Luce}}, \bibinfo {author} {\bibfnamefont
  {W.}~\bibnamefont {Fan}}, \bibinfo {author} {\bibfnamefont {K.}~\bibnamefont
  {Liu}}, \bibinfo {author} {\bibfnamefont {J.}~\bibnamefont {Zhou}}, \bibinfo
  {author} {\bibfnamefont {Y.-S.}\ \bibnamefont {Huang}}, \bibinfo {author}
  {\bibfnamefont {C.-H.}\ \bibnamefont {Ho}}, \bibinfo {author} {\bibfnamefont
  {J.}~\bibnamefont {Yan}}, \bibinfo {author} {\bibfnamefont {D.~F.}\
  \bibnamefont {Ogletree}}, \bibinfo {author} {\bibfnamefont {S.}~\bibnamefont
  {Aloni}}, \bibinfo {author} {\bibfnamefont {J.}~\bibnamefont {Ji}}, \bibinfo
  {author} {\bibfnamefont {S.}~\bibnamefont {Li}}, \bibinfo {author}
  {\bibfnamefont {J.}~\bibnamefont {Li}}, \bibinfo {author} {\bibfnamefont
  {F.~M.}\ \bibnamefont {Peeters}},\ and\ \bibinfo {author} {\bibfnamefont
  {J.}~\bibnamefont {Wu}},\ }\href {https://doi.org/10.1038/ncomms4252}
  {\bibfield  {journal} {\bibinfo  {journal} {Nature Communications}\ }\textbf
  {\bibinfo {volume} {5}},\ \bibinfo {pages} {3252} (\bibinfo {year}
  {2014})}\BibitemShut {NoStop}%
\bibitem [{\citenamefont {Yu}\ \emph {et~al.}(2015)\citenamefont {Yu},
  \citenamefont {Cai},\ and\ \citenamefont {Zhang}}]{Yu2015}%
  \BibitemOpen
  \bibfield  {author} {\bibinfo {author} {\bibfnamefont {Z.~G.}\ \bibnamefont
  {Yu}}, \bibinfo {author} {\bibfnamefont {Y.}~\bibnamefont {Cai}},\ and\
  \bibinfo {author} {\bibfnamefont {Y.-W.}\ \bibnamefont {Zhang}},\ }\href
  {https://doi.org/10.1038/srep13783} {\bibfield  {journal} {\bibinfo
  {journal} {Scientific Reports}\ }\textbf {\bibinfo {volume} {5}},\ \bibinfo
  {pages} {13783} (\bibinfo {year} {2015})}\BibitemShut {NoStop}%
\bibitem [{\citenamefont {Jariwala}\ \emph {et~al.}(2016)\citenamefont
  {Jariwala}, \citenamefont {Voiry}, \citenamefont {Jindal}, \citenamefont
  {Chalke}, \citenamefont {Bapat}, \citenamefont {Thamizhavel}, \citenamefont
  {Chhowalla}, \citenamefont {Deshmukh},\ and\ \citenamefont
  {Bhattacharya}}]{Jariwala2016}%
  \BibitemOpen
  \bibfield  {author} {\bibinfo {author} {\bibfnamefont {B.}~\bibnamefont
  {Jariwala}}, \bibinfo {author} {\bibfnamefont {D.}~\bibnamefont {Voiry}},
  \bibinfo {author} {\bibfnamefont {A.}~\bibnamefont {Jindal}}, \bibinfo
  {author} {\bibfnamefont {B.~A.}\ \bibnamefont {Chalke}}, \bibinfo {author}
  {\bibfnamefont {R.}~\bibnamefont {Bapat}}, \bibinfo {author} {\bibfnamefont
  {A.}~\bibnamefont {Thamizhavel}}, \bibinfo {author} {\bibfnamefont
  {M.}~\bibnamefont {Chhowalla}}, \bibinfo {author} {\bibfnamefont
  {M.}~\bibnamefont {Deshmukh}},\ and\ \bibinfo {author} {\bibfnamefont
  {A.}~\bibnamefont {Bhattacharya}},\ }\href
  {https://doi.org/10.1021/acs.chemmater.6b00364} {\bibfield  {journal}
  {\bibinfo  {journal} {Chemistry of Materials}\ }\textbf {\bibinfo {volume}
  {28}},\ \bibinfo {pages} {3352} (\bibinfo {year} {2016})}\BibitemShut
  {NoStop}%
\bibitem [{\citenamefont {Echeverry}\ and\ \citenamefont
  {Gerber}(2018)}]{Echeverry2018}%
  \BibitemOpen
  \bibfield  {author} {\bibinfo {author} {\bibfnamefont {J.~P.}\ \bibnamefont
  {Echeverry}}\ and\ \bibinfo {author} {\bibfnamefont {I.~C.}\ \bibnamefont
  {Gerber}},\ }\href {https://doi.org/10.1103/PhysRevB.97.075123} {\bibfield
  {journal} {\bibinfo  {journal} {Phys. Rev. B}\ }\textbf {\bibinfo {volume}
  {97}},\ \bibinfo {pages} {075123} (\bibinfo {year} {2018})}\BibitemShut
  {NoStop}%
\bibitem [{\citenamefont {Sim}\ \emph {et~al.}(2019)\citenamefont {Sim},
  \citenamefont {Lee}, \citenamefont {Lee}, \citenamefont {Bae}, \citenamefont
  {Noh}, \citenamefont {Cha}, \citenamefont {Jo}, \citenamefont {Lee},\ and\
  \citenamefont {Choi}}]{Sim2019}%
  \BibitemOpen
  \bibfield  {author} {\bibinfo {author} {\bibfnamefont {S.}~\bibnamefont
  {Sim}}, \bibinfo {author} {\bibfnamefont {D.}~\bibnamefont {Lee}}, \bibinfo
  {author} {\bibfnamefont {J.}~\bibnamefont {Lee}}, \bibinfo {author}
  {\bibfnamefont {H.}~\bibnamefont {Bae}}, \bibinfo {author} {\bibfnamefont
  {M.}~\bibnamefont {Noh}}, \bibinfo {author} {\bibfnamefont {S.}~\bibnamefont
  {Cha}}, \bibinfo {author} {\bibfnamefont {M.-H.}\ \bibnamefont {Jo}},
  \bibinfo {author} {\bibfnamefont {K.}~\bibnamefont {Lee}},\ and\ \bibinfo
  {author} {\bibfnamefont {H.}~\bibnamefont {Choi}},\ }\href
  {https://doi.org/10.1021/acs.nanolett.9b03173} {\bibfield  {journal}
  {\bibinfo  {journal} {Nano Letters}\ }\textbf {\bibinfo {volume} {19}},\
  \bibinfo {pages} {7464} (\bibinfo {year} {2019})}\BibitemShut {NoStop}%
\bibitem [{\citenamefont {Wang}\ \emph {et~al.}(2020)\citenamefont {Wang},
  \citenamefont {Zhou}, \citenamefont {Xiang}, \citenamefont {Ng},
  \citenamefont {Watanabe}, \citenamefont {Taniguchi},\ and\ \citenamefont
  {Eda}}]{WangJunyong2020}%
  \BibitemOpen
  \bibfield  {author} {\bibinfo {author} {\bibfnamefont {J.}~\bibnamefont
  {Wang}}, \bibinfo {author} {\bibfnamefont {Y.~J.}\ \bibnamefont {Zhou}},
  \bibinfo {author} {\bibfnamefont {D.}~\bibnamefont {Xiang}}, \bibinfo
  {author} {\bibfnamefont {S.~J.}\ \bibnamefont {Ng}}, \bibinfo {author}
  {\bibfnamefont {K.}~\bibnamefont {Watanabe}}, \bibinfo {author}
  {\bibfnamefont {T.}~\bibnamefont {Taniguchi}},\ and\ \bibinfo {author}
  {\bibfnamefont {G.}~\bibnamefont {Eda}},\ }\href
  {https://doi.org/https://doi.org/10.1002/adma.202001890} {\bibfield
  {journal} {\bibinfo  {journal} {Advanced Materials}\ }\textbf {\bibinfo
  {volume} {32}},\ \bibinfo {pages} {2001890} (\bibinfo {year}
  {2020})}\BibitemShut {NoStop}%
\bibitem [{\citenamefont {Zhang}\ \emph {et~al.}(2015)\citenamefont {Zhang},
  \citenamefont {Jin}, \citenamefont {Yuan}, \citenamefont {Wang},
  \citenamefont {Zhang}, \citenamefont {Tang}, \citenamefont {Liu},
  \citenamefont {Zhou}, \citenamefont {Hu},\ and\ \citenamefont
  {Xiu}}]{ZhangEnze2015}%
  \BibitemOpen
  \bibfield  {author} {\bibinfo {author} {\bibfnamefont {E.}~\bibnamefont
  {Zhang}}, \bibinfo {author} {\bibfnamefont {Y.}~\bibnamefont {Jin}}, \bibinfo
  {author} {\bibfnamefont {X.}~\bibnamefont {Yuan}}, \bibinfo {author}
  {\bibfnamefont {W.}~\bibnamefont {Wang}}, \bibinfo {author} {\bibfnamefont
  {C.}~\bibnamefont {Zhang}}, \bibinfo {author} {\bibfnamefont
  {L.}~\bibnamefont {Tang}}, \bibinfo {author} {\bibfnamefont {S.}~\bibnamefont
  {Liu}}, \bibinfo {author} {\bibfnamefont {P.}~\bibnamefont {Zhou}}, \bibinfo
  {author} {\bibfnamefont {W.}~\bibnamefont {Hu}},\ and\ \bibinfo {author}
  {\bibfnamefont {F.}~\bibnamefont {Xiu}},\ }\href
  {https://doi.org/https://doi.org/10.1002/adfm.201500969} {\bibfield
  {journal} {\bibinfo  {journal} {Advanced Functional Materials}\ }\textbf
  {\bibinfo {volume} {25}},\ \bibinfo {pages} {4076} (\bibinfo {year}
  {2015})}\BibitemShut {NoStop}%
\bibitem [{\citenamefont {Kwon}\ \emph {et~al.}(2019)\citenamefont {Kwon},
  \citenamefont {Shin}, \citenamefont {Kwon}, \citenamefont {Lee},
  \citenamefont {Park}, \citenamefont {Watanabe}, \citenamefont {Taniguchi},
  \citenamefont {Kim}, \citenamefont {Lee}, \citenamefont {Im},\ and\
  \citenamefont {Lee}}]{Kwon2019}%
  \BibitemOpen
  \bibfield  {author} {\bibinfo {author} {\bibfnamefont {J.}~\bibnamefont
  {Kwon}}, \bibinfo {author} {\bibfnamefont {Y.}~\bibnamefont {Shin}}, \bibinfo
  {author} {\bibfnamefont {H.}~\bibnamefont {Kwon}}, \bibinfo {author}
  {\bibfnamefont {J.~Y.}\ \bibnamefont {Lee}}, \bibinfo {author} {\bibfnamefont
  {H.}~\bibnamefont {Park}}, \bibinfo {author} {\bibfnamefont {K.}~\bibnamefont
  {Watanabe}}, \bibinfo {author} {\bibfnamefont {T.}~\bibnamefont {Taniguchi}},
  \bibinfo {author} {\bibfnamefont {J.}~\bibnamefont {Kim}}, \bibinfo {author}
  {\bibfnamefont {C.-H.}\ \bibnamefont {Lee}}, \bibinfo {author} {\bibfnamefont
  {S.}~\bibnamefont {Im}},\ and\ \bibinfo {author} {\bibfnamefont {G.-H.}\
  \bibnamefont {Lee}},\ }\href {https://doi.org/10.1038/s41598-019-46730-7}
  {\bibfield  {journal} {\bibinfo  {journal} {Scientific Reports}\ }\textbf
  {\bibinfo {volume} {9}},\ \bibinfo {pages} {10354} (\bibinfo {year}
  {2019})}\BibitemShut {NoStop}%
\bibitem [{\citenamefont {Aslan}\ \emph {et~al.}(2016)\citenamefont {Aslan},
  \citenamefont {Chenet}, \citenamefont {van~der Zande}, \citenamefont {Hone},\
  and\ \citenamefont {Heinz}}]{Aslan2016}%
  \BibitemOpen
  \bibfield  {author} {\bibinfo {author} {\bibfnamefont {O.~B.}\ \bibnamefont
  {Aslan}}, \bibinfo {author} {\bibfnamefont {D.~A.}\ \bibnamefont {Chenet}},
  \bibinfo {author} {\bibfnamefont {A.~M.}\ \bibnamefont {van~der Zande}},
  \bibinfo {author} {\bibfnamefont {J.~C.}\ \bibnamefont {Hone}},\ and\
  \bibinfo {author} {\bibfnamefont {T.~F.}\ \bibnamefont {Heinz}},\ }\href
  {https://doi.org/10.1021/acsphotonics.5b00486} {\bibfield  {journal}
  {\bibinfo  {journal} {ACS Photonics}\ }\textbf {\bibinfo {volume} {3}},\
  \bibinfo {pages} {96} (\bibinfo {year} {2016})}\BibitemShut {NoStop}%
\bibitem [{\citenamefont {Jadczak}\ \emph {et~al.}(2019)\citenamefont
  {Jadczak}, \citenamefont {Kutrowska-Girzycka}, \citenamefont {Smolenski},
  \citenamefont {Kossacki}, \citenamefont {Huang},\ and\ \citenamefont
  {Bryja}}]{Jadczak2019SR}%
  \BibitemOpen
  \bibfield  {author} {\bibinfo {author} {\bibfnamefont {J.}~\bibnamefont
  {Jadczak}}, \bibinfo {author} {\bibfnamefont {J.}~\bibnamefont
  {Kutrowska-Girzycka}}, \bibinfo {author} {\bibfnamefont {T.}~\bibnamefont
  {Smolenski}}, \bibinfo {author} {\bibfnamefont {P.}~\bibnamefont {Kossacki}},
  \bibinfo {author} {\bibfnamefont {Y.~S.}\ \bibnamefont {Huang}},\ and\
  \bibinfo {author} {\bibfnamefont {L.}~\bibnamefont {Bryja}},\ }\href
  {https://doi.org/10.1038/s41598-018-37655-8} {\bibfield  {journal} {\bibinfo
  {journal} {Scientific Reports}\ }\textbf {\bibinfo {volume} {9}},\ \bibinfo
  {pages} {1578} (\bibinfo {year} {2019})}\BibitemShut {NoStop}%
\bibitem [{\citenamefont {Urban}\ \emph {et~al.}(2018)\citenamefont {Urban},
  \citenamefont {Baranowski}, \citenamefont {Kuc}, \citenamefont
  {K{\l}opotowski}, \citenamefont {Surrente}, \citenamefont {Ma}, \citenamefont
  {W{\l}odarczyk}, \citenamefont {Suchocki}, \citenamefont {Ovchinnikov},
  \citenamefont {Heine}, \citenamefont {Maude}, \citenamefont {Kis},\ and\
  \citenamefont {Plochocka}}]{Urban_2018}%
  \BibitemOpen
  \bibfield  {author} {\bibinfo {author} {\bibfnamefont {J.~M.}\ \bibnamefont
  {Urban}}, \bibinfo {author} {\bibfnamefont {M.}~\bibnamefont {Baranowski}},
  \bibinfo {author} {\bibfnamefont {A.}~\bibnamefont {Kuc}}, \bibinfo {author}
  {\bibfnamefont {{\L}.}~\bibnamefont {K{\l}opotowski}}, \bibinfo {author}
  {\bibfnamefont {A.}~\bibnamefont {Surrente}}, \bibinfo {author}
  {\bibfnamefont {Y.}~\bibnamefont {Ma}}, \bibinfo {author} {\bibfnamefont
  {D.}~\bibnamefont {W{\l}odarczyk}}, \bibinfo {author} {\bibfnamefont
  {A.}~\bibnamefont {Suchocki}}, \bibinfo {author} {\bibfnamefont
  {D.}~\bibnamefont {Ovchinnikov}}, \bibinfo {author} {\bibfnamefont
  {T.}~\bibnamefont {Heine}}, \bibinfo {author} {\bibfnamefont {D.~K.}\
  \bibnamefont {Maude}}, \bibinfo {author} {\bibfnamefont {A.}~\bibnamefont
  {Kis}},\ and\ \bibinfo {author} {\bibfnamefont {P.}~\bibnamefont
  {Plochocka}},\ }\href {https://doi.org/10.1088/2053-1583/aae9b9} {\bibfield
  {journal} {\bibinfo  {journal} {2D Materials}\ }\textbf {\bibinfo {volume}
  {6}},\ \bibinfo {pages} {015012} (\bibinfo {year} {2018})}\BibitemShut
  {NoStop}%
\bibitem [{\citenamefont {Sim}\ \emph {et~al.}(2018)\citenamefont {Sim},
  \citenamefont {Lee}, \citenamefont {Trifonov}, \citenamefont {Kim},
  \citenamefont {Cha}, \citenamefont {Sung}, \citenamefont {Cho}, \citenamefont
  {Shim}, \citenamefont {Jo},\ and\ \citenamefont {Choi}}]{Sim2018}%
  \BibitemOpen
  \bibfield  {author} {\bibinfo {author} {\bibfnamefont {S.}~\bibnamefont
  {Sim}}, \bibinfo {author} {\bibfnamefont {D.}~\bibnamefont {Lee}}, \bibinfo
  {author} {\bibfnamefont {A.~V.}\ \bibnamefont {Trifonov}}, \bibinfo {author}
  {\bibfnamefont {T.}~\bibnamefont {Kim}}, \bibinfo {author} {\bibfnamefont
  {S.}~\bibnamefont {Cha}}, \bibinfo {author} {\bibfnamefont {J.~H.}\
  \bibnamefont {Sung}}, \bibinfo {author} {\bibfnamefont {S.}~\bibnamefont
  {Cho}}, \bibinfo {author} {\bibfnamefont {W.}~\bibnamefont {Shim}}, \bibinfo
  {author} {\bibfnamefont {M.-H.}\ \bibnamefont {Jo}},\ and\ \bibinfo {author}
  {\bibfnamefont {H.}~\bibnamefont {Choi}},\ }\href
  {https://doi.org/10.1038/s41467-017-02802-8} {\bibfield  {journal} {\bibinfo
  {journal} {Nature Communications}\ }\textbf {\bibinfo {volume} {9}},\
  \bibinfo {pages} {351} (\bibinfo {year} {2018})}\BibitemShut {NoStop}%
\bibitem [{\citenamefont {Wang}\ \emph {et~al.}(2019)\citenamefont {Wang},
  \citenamefont {Shinokita}, \citenamefont {Lim}, \citenamefont {Mohamed},
  \citenamefont {Miyauchi}, \citenamefont {Cuong}, \citenamefont {Okada},\ and\
  \citenamefont {Matsuda}}]{Xiaofan2019}%
  \BibitemOpen
  \bibfield  {author} {\bibinfo {author} {\bibfnamefont {X.}~\bibnamefont
  {Wang}}, \bibinfo {author} {\bibfnamefont {K.}~\bibnamefont {Shinokita}},
  \bibinfo {author} {\bibfnamefont {H.~E.}\ \bibnamefont {Lim}}, \bibinfo
  {author} {\bibfnamefont {N.~B.}\ \bibnamefont {Mohamed}}, \bibinfo {author}
  {\bibfnamefont {Y.}~\bibnamefont {Miyauchi}}, \bibinfo {author}
  {\bibfnamefont {N.~T.}\ \bibnamefont {Cuong}}, \bibinfo {author}
  {\bibfnamefont {S.}~\bibnamefont {Okada}},\ and\ \bibinfo {author}
  {\bibfnamefont {K.}~\bibnamefont {Matsuda}},\ }\href
  {https://doi.org/10.1002/adfm.201806169} {\bibfield  {journal} {\bibinfo
  {journal} {Advanced Functional Materials}\ }\textbf {\bibinfo {volume}
  {29}},\ \bibinfo {pages} {1806169} (\bibinfo {year} {2019})}\BibitemShut
  {NoStop}%
\bibitem [{\citenamefont {Ho}\ and\ \citenamefont {Liu}(2019)}]{HO2019641}%
  \BibitemOpen
  \bibfield  {author} {\bibinfo {author} {\bibfnamefont {C.-H.}\ \bibnamefont
  {Ho}}\ and\ \bibinfo {author} {\bibfnamefont {Z.-Z.}\ \bibnamefont {Liu}},\
  }\href {https://doi.org/https://doi.org/10.1016/j.nanoen.2018.12.014}
  {\bibfield  {journal} {\bibinfo  {journal} {Nano Energy}\ }\textbf {\bibinfo
  {volume} {56}},\ \bibinfo {pages} {641 } (\bibinfo {year}
  {2019})}\BibitemShut {NoStop}%
\bibitem [{\citenamefont {Kipczak}\ \emph {et~al.}(2020)\citenamefont
  {Kipczak}, \citenamefont {Grzeszczyk}, \citenamefont {Olkowska-Pucko},
  \citenamefont {Babiński},\ and\ \citenamefont {Molas}}]{kipczak2020optical}%
  \BibitemOpen
  \bibfield  {author} {\bibinfo {author} {\bibfnamefont {{\L}.}~\bibnamefont
  {Kipczak}}, \bibinfo {author} {\bibfnamefont {M.}~\bibnamefont {Grzeszczyk}},
  \bibinfo {author} {\bibfnamefont {K.}~\bibnamefont {Olkowska-Pucko}},
  \bibinfo {author} {\bibfnamefont {A.}~\bibnamefont {Babiński}},\ and\
  \bibinfo {author} {\bibfnamefont {M.~R.}\ \bibnamefont {Molas}},\ }\href
  {https://doi.org/10.1063/5.0015289} {\bibfield  {journal} {\bibinfo
  {journal} {Journal of Applied Physics}\ }\textbf {\bibinfo {volume} {128}},\
  \bibinfo {pages} {044302} (\bibinfo {year} {2020})}\BibitemShut {NoStop}%
\bibitem [{\citenamefont {Dhara}\ \emph {et~al.}(2020)\citenamefont {Dhara},
  \citenamefont {Chakrabarty}, \citenamefont {Das}, \citenamefont {Pattanayak},
  \citenamefont {Paul}, \citenamefont {Mukherjee},\ and\ \citenamefont
  {Dhara}}]{Dhara2020AdditionalEF}%
  \BibitemOpen
  \bibfield  {author} {\bibinfo {author} {\bibfnamefont {A.}~\bibnamefont
  {Dhara}}, \bibinfo {author} {\bibfnamefont {D.}~\bibnamefont {Chakrabarty}},
  \bibinfo {author} {\bibfnamefont {P.}~\bibnamefont {Das}}, \bibinfo {author}
  {\bibfnamefont {A.~K.}\ \bibnamefont {Pattanayak}}, \bibinfo {author}
  {\bibfnamefont {S.}~\bibnamefont {Paul}}, \bibinfo {author} {\bibfnamefont
  {S.}~\bibnamefont {Mukherjee}},\ and\ \bibinfo {author} {\bibfnamefont
  {S.}~\bibnamefont {Dhara}},\ }\href
  {https://doi.org/10.1103/PhysRevB.102.161404} {\bibfield  {journal} {\bibinfo
   {journal} {Phys. Rev. B}\ }\textbf {\bibinfo {volume} {102}},\ \bibinfo
  {pages} {161404} (\bibinfo {year} {2020})}\BibitemShut {NoStop}%
\bibitem [{\citenamefont {Arora}\ \emph {et~al.}(2017)\citenamefont {Arora},
  \citenamefont {Noky}, \citenamefont {Dr{\"u}ppel}, \citenamefont {Jariwala},
  \citenamefont {Deilmann}, \citenamefont {Schneider}, \citenamefont {Schmidt},
  \citenamefont {Del Pozo-Zamudio}, \citenamefont {Stiehm}, \citenamefont
  {Bhattacharya}, \citenamefont {Kr{\"u}ger}, \citenamefont {Michaelis~de
  Vasconcellos}, \citenamefont {Rohlfing},\ and\ \citenamefont
  {Bratschitsch}}]{Arora2017}%
  \BibitemOpen
  \bibfield  {author} {\bibinfo {author} {\bibfnamefont {A.}~\bibnamefont
  {Arora}}, \bibinfo {author} {\bibfnamefont {J.}~\bibnamefont {Noky}},
  \bibinfo {author} {\bibfnamefont {M.}~\bibnamefont {Dr{\"u}ppel}}, \bibinfo
  {author} {\bibfnamefont {B.}~\bibnamefont {Jariwala}}, \bibinfo {author}
  {\bibfnamefont {T.}~\bibnamefont {Deilmann}}, \bibinfo {author}
  {\bibfnamefont {R.}~\bibnamefont {Schneider}}, \bibinfo {author}
  {\bibfnamefont {R.}~\bibnamefont {Schmidt}}, \bibinfo {author} {\bibfnamefont
  {O.}~\bibnamefont {Del Pozo-Zamudio}}, \bibinfo {author} {\bibfnamefont
  {T.}~\bibnamefont {Stiehm}}, \bibinfo {author} {\bibfnamefont
  {A.}~\bibnamefont {Bhattacharya}}, \bibinfo {author} {\bibfnamefont
  {P.}~\bibnamefont {Kr{\"u}ger}}, \bibinfo {author} {\bibfnamefont
  {S.}~\bibnamefont {Michaelis~de Vasconcellos}}, \bibinfo {author}
  {\bibfnamefont {M.}~\bibnamefont {Rohlfing}},\ and\ \bibinfo {author}
  {\bibfnamefont {R.}~\bibnamefont {Bratschitsch}},\ }\href
  {https://doi.org/10.1021/acs.nanolett.7b00765} {\bibfield  {journal}
  {\bibinfo  {journal} {Nano Letters}\ }\textbf {\bibinfo {volume} {17}},\
  \bibinfo {pages} {3202} (\bibinfo {year} {2017})}\BibitemShut {NoStop}%
\bibitem [{\citenamefont {Guti{\'{e}}rrez-Lezama}\ \emph
  {et~al.}(2016)\citenamefont {Guti{\'{e}}rrez-Lezama}, \citenamefont {Reddy},
  \citenamefont {Ubrig},\ and\ \citenamefont
  {Morpurgo}}]{Guti_rrez_Lezama_2016}%
  \BibitemOpen
  \bibfield  {author} {\bibinfo {author} {\bibfnamefont {I.}~\bibnamefont
  {Guti{\'{e}}rrez-Lezama}}, \bibinfo {author} {\bibfnamefont {B.~A.}\
  \bibnamefont {Reddy}}, \bibinfo {author} {\bibfnamefont {N.}~\bibnamefont
  {Ubrig}},\ and\ \bibinfo {author} {\bibfnamefont {A.~F.}\ \bibnamefont
  {Morpurgo}},\ }\href {https://doi.org/10.1088/2053-1583/3/4/045016}
  {\bibfield  {journal} {\bibinfo  {journal} {2D Materials}\ }\textbf {\bibinfo
  {volume} {3}},\ \bibinfo {pages} {045016} (\bibinfo {year}
  {2016})}\BibitemShut {NoStop}%
\bibitem [{\citenamefont {Ho}\ \emph {et~al.}(1998{\natexlab{b}})\citenamefont
  {Ho}, \citenamefont {Huang}, \citenamefont {Tiong},\ and\ \citenamefont
  {Liao}}]{Ho_1998}%
  \BibitemOpen
  \bibfield  {author} {\bibinfo {author} {\bibfnamefont {C.~H.}\ \bibnamefont
  {Ho}}, \bibinfo {author} {\bibfnamefont {Y.~S.}\ \bibnamefont {Huang}},
  \bibinfo {author} {\bibfnamefont {K.~K.}\ \bibnamefont {Tiong}},\ and\
  \bibinfo {author} {\bibfnamefont {P.~C.}\ \bibnamefont {Liao}},\ }\href
  {https://doi.org/10.1103/PhysRevB.58.16130} {\bibfield  {journal} {\bibinfo
  {journal} {Phys. Rev. B}\ }\textbf {\bibinfo {volume} {58}},\ \bibinfo
  {pages} {16130} (\bibinfo {year} {1998}{\natexlab{b}})}\BibitemShut {NoStop}%
\bibitem [{\citenamefont {Ho}\ \emph {et~al.}(1999)\citenamefont {Ho},
  \citenamefont {Huang}, \citenamefont {Chen}, \citenamefont {Dann},\ and\
  \citenamefont {Tiong}}]{Ho_1999}%
  \BibitemOpen
  \bibfield  {author} {\bibinfo {author} {\bibfnamefont {C.~H.}\ \bibnamefont
  {Ho}}, \bibinfo {author} {\bibfnamefont {Y.~S.}\ \bibnamefont {Huang}},
  \bibinfo {author} {\bibfnamefont {J.~L.}\ \bibnamefont {Chen}}, \bibinfo
  {author} {\bibfnamefont {T.~E.}\ \bibnamefont {Dann}},\ and\ \bibinfo
  {author} {\bibfnamefont {K.~K.}\ \bibnamefont {Tiong}},\ }\href
  {https://doi.org/10.1103/PhysRevB.60.15766} {\bibfield  {journal} {\bibinfo
  {journal} {Phys. Rev. B}\ }\textbf {\bibinfo {volume} {60}},\ \bibinfo
  {pages} {15766} (\bibinfo {year} {1999})}\BibitemShut {NoStop}%
\bibitem [{\citenamefont {Ho}\ \emph {et~al.}(2004)\citenamefont {Ho},
  \citenamefont {Lee},\ and\ \citenamefont {Wu}}]{Ho_2004}%
  \BibitemOpen
  \bibfield  {author} {\bibinfo {author} {\bibfnamefont {C.~H.}\ \bibnamefont
  {Ho}}, \bibinfo {author} {\bibfnamefont {H.~W.}\ \bibnamefont {Lee}},\ and\
  \bibinfo {author} {\bibfnamefont {C.~C.}\ \bibnamefont {Wu}},\ }\href
  {https://doi.org/10.1088/0953-8984/16/32/026} {\bibfield  {journal} {\bibinfo
   {journal} {Journal of Physics: Condensed Matter}\ }\textbf {\bibinfo
  {volume} {16}},\ \bibinfo {pages} {5937} (\bibinfo {year}
  {2004})}\BibitemShut {NoStop}%
\bibitem [{\citenamefont {Zelewski}\ and\ \citenamefont
  {Kudrawiec}(2017)}]{Zelewski2017}%
  \BibitemOpen
  \bibfield  {author} {\bibinfo {author} {\bibfnamefont {S.~J.}\ \bibnamefont
  {Zelewski}}\ and\ \bibinfo {author} {\bibfnamefont {R.}~\bibnamefont
  {Kudrawiec}},\ }\href {https://doi.org/10.1038/s41598-017-15763-1} {\bibfield
   {journal} {\bibinfo  {journal} {Scientific Reports}\ }\textbf {\bibinfo
  {volume} {7}},\ \bibinfo {pages} {15365} (\bibinfo {year}
  {2017})}\BibitemShut {NoStop}%
\bibitem [{\citenamefont {Gunasekera}\ \emph {et~al.}(2018)\citenamefont
  {Gunasekera}, \citenamefont {Wolverson}, \citenamefont {Hart},\ and\
  \citenamefont {Mucha-Kruczynski}}]{Gunasekera2018}%
  \BibitemOpen
  \bibfield  {author} {\bibinfo {author} {\bibfnamefont {S.~M.}\ \bibnamefont
  {Gunasekera}}, \bibinfo {author} {\bibfnamefont {D.}~\bibnamefont
  {Wolverson}}, \bibinfo {author} {\bibfnamefont {L.~S.}\ \bibnamefont
  {Hart}},\ and\ \bibinfo {author} {\bibfnamefont {M.}~\bibnamefont
  {Mucha-Kruczynski}},\ }\href {https://doi.org/10.1007/s11664-018-6239-0}
  {\bibfield  {journal} {\bibinfo  {journal} {Journal of Electronic Materials}\
  }\textbf {\bibinfo {volume} {47}},\ \bibinfo {pages} {4314} (\bibinfo {year}
  {2018})}\BibitemShut {NoStop}%
\bibitem [{\citenamefont {He}(1991)}]{He1991}%
  \BibitemOpen
  \bibfield  {author} {\bibinfo {author} {\bibfnamefont {X.-F.}\ \bibnamefont
  {He}},\ }\href {https://doi.org/10.1103/PhysRevB.43.2063} {\bibfield
  {journal} {\bibinfo  {journal} {Phys. Rev. B}\ }\textbf {\bibinfo {volume}
  {43}},\ \bibinfo {pages} {2063} (\bibinfo {year} {1991})}\BibitemShut
  {NoStop}%
\bibitem [{\citenamefont {Mathieu}\ \emph
  {et~al.}(1992{\natexlab{a}})\citenamefont {Mathieu}, \citenamefont
  {Lefebvre},\ and\ \citenamefont {Christol}}]{Mathieu1992}%
  \BibitemOpen
  \bibfield  {author} {\bibinfo {author} {\bibfnamefont {H.}~\bibnamefont
  {Mathieu}}, \bibinfo {author} {\bibfnamefont {P.}~\bibnamefont {Lefebvre}},\
  and\ \bibinfo {author} {\bibfnamefont {P.}~\bibnamefont {Christol}},\ }\href
  {https://doi.org/10.1063/1.352137} {\bibfield  {journal} {\bibinfo  {journal}
  {Journal of Applied Physics}\ }\textbf {\bibinfo {volume} {72}},\ \bibinfo
  {pages} {300} (\bibinfo {year} {1992}{\natexlab{a}})},\ \Eprint
  {https://arxiv.org/abs/https://doi.org/10.1063/1.352137}
  {https://doi.org/10.1063/1.352137} \BibitemShut {NoStop}%
\bibitem [{\citenamefont {Mathieu}\ \emph
  {et~al.}(1992{\natexlab{b}})\citenamefont {Mathieu}, \citenamefont
  {Lefebvre},\ and\ \citenamefont {Christol}}]{PhysRevB.46.4092}%
  \BibitemOpen
  \bibfield  {author} {\bibinfo {author} {\bibfnamefont {H.}~\bibnamefont
  {Mathieu}}, \bibinfo {author} {\bibfnamefont {P.}~\bibnamefont {Lefebvre}},\
  and\ \bibinfo {author} {\bibfnamefont {P.}~\bibnamefont {Christol}},\ }\href
  {https://doi.org/10.1103/PhysRevB.46.4092} {\bibfield  {journal} {\bibinfo
  {journal} {Phys. Rev. B}\ }\textbf {\bibinfo {volume} {46}},\ \bibinfo
  {pages} {4092} (\bibinfo {year} {1992}{\natexlab{b}})}\BibitemShut {NoStop}%
\bibitem [{\citenamefont {Christol}\ \emph {et~al.}(1993)\citenamefont
  {Christol}, \citenamefont {Lefebvre},\ and\ \citenamefont
  {Mathieu}}]{Christol1993}%
  \BibitemOpen
  \bibfield  {author} {\bibinfo {author} {\bibfnamefont {P.}~\bibnamefont
  {Christol}}, \bibinfo {author} {\bibfnamefont {P.}~\bibnamefont {Lefebvre}},\
  and\ \bibinfo {author} {\bibfnamefont {H.}~\bibnamefont {Mathieu}},\ }\href
  {https://doi.org/10.1063/1.354224} {\bibfield  {journal} {\bibinfo  {journal}
  {Journal of Applied Physics}\ }\textbf {\bibinfo {volume} {74}},\ \bibinfo
  {pages} {5626} (\bibinfo {year} {1993})},\ \Eprint
  {https://arxiv.org/abs/https://doi.org/10.1063/1.354224}
  {https://doi.org/10.1063/1.354224} \BibitemShut {NoStop}%
\bibitem [{\citenamefont {Molas}\ \emph {et~al.}(2019)\citenamefont {Molas},
  \citenamefont {Slobodeniuk}, \citenamefont {Nogajewski}, \citenamefont
  {Bartos}, \citenamefont {Bala}, \citenamefont {Babi\ifmmode~\acute{n}\else
  \'{n}\fi{}ski}, \citenamefont {Watanabe}, \citenamefont {Taniguchi},
  \citenamefont {Faugeras},\ and\ \citenamefont {Potemski}}]{Molas5s}%
  \BibitemOpen
  \bibfield  {author} {\bibinfo {author} {\bibfnamefont {M.~R.}\ \bibnamefont
  {Molas}}, \bibinfo {author} {\bibfnamefont {A.~O.}\ \bibnamefont
  {Slobodeniuk}}, \bibinfo {author} {\bibfnamefont {K.}~\bibnamefont
  {Nogajewski}}, \bibinfo {author} {\bibfnamefont {M.}~\bibnamefont {Bartos}},
  \bibinfo {author} {\bibfnamefont {L.}~\bibnamefont {Bala}}, \bibinfo {author}
  {\bibfnamefont {A.}~\bibnamefont {Babi\ifmmode~\acute{n}\else
  \'{n}\fi{}ski}}, \bibinfo {author} {\bibfnamefont {K.}~\bibnamefont
  {Watanabe}}, \bibinfo {author} {\bibfnamefont {T.}~\bibnamefont {Taniguchi}},
  \bibinfo {author} {\bibfnamefont {C.}~\bibnamefont {Faugeras}},\ and\
  \bibinfo {author} {\bibfnamefont {M.}~\bibnamefont {Potemski}},\ }\href
  {https://doi.org/10.1103/PhysRevLett.123.136801} {\bibfield  {journal}
  {\bibinfo  {journal} {Phys. Rev. Lett.}\ }\textbf {\bibinfo {volume} {123}},\
  \bibinfo {pages} {136801} (\bibinfo {year} {2019})}\BibitemShut {NoStop}%
\bibitem [{\citenamefont {Molas}\ \emph {et~al.}(2017)\citenamefont {Molas},
  \citenamefont {Faugeras}, \citenamefont {Slobodeniuk}, \citenamefont
  {Nogajewski}, \citenamefont {Bartos}, \citenamefont {Basko},\ and\
  \citenamefont {Potemski}}]{molas}%
  \BibitemOpen
  \bibfield  {author} {\bibinfo {author} {\bibfnamefont {M.~R.}\ \bibnamefont
  {Molas}}, \bibinfo {author} {\bibfnamefont {C.}~\bibnamefont {Faugeras}},
  \bibinfo {author} {\bibfnamefont {A.~O.}\ \bibnamefont {Slobodeniuk}},
  \bibinfo {author} {\bibfnamefont {K.}~\bibnamefont {Nogajewski}}, \bibinfo
  {author} {\bibfnamefont {M.}~\bibnamefont {Bartos}}, \bibinfo {author}
  {\bibfnamefont {D.~M.}\ \bibnamefont {Basko}},\ and\ \bibinfo {author}
  {\bibfnamefont {M.}~\bibnamefont {Potemski}},\ }\href
  {https://doi.org/10.1088/2053-1583/aa5521} {\bibfield  {journal} {\bibinfo
  {journal} {2D Materials}\ }\textbf {\bibinfo {volume} {4}},\ \bibinfo {pages}
  {021003} (\bibinfo {year} {2017})}\BibitemShut {NoStop}%
\bibitem [{\citenamefont {Wang}\ \emph {et~al.}(2017)\citenamefont {Wang},
  \citenamefont {Robert}, \citenamefont {Glazov}, \citenamefont {Cadiz},
  \citenamefont {Courtade}, \citenamefont {Amand}, \citenamefont {Lagarde},
  \citenamefont {Taniguchi}, \citenamefont {Watanabe}, \citenamefont
  {Urbaszek},\ and\ \citenamefont {Marie}}]{wang2017}%
  \BibitemOpen
  \bibfield  {author} {\bibinfo {author} {\bibfnamefont {G.}~\bibnamefont
  {Wang}}, \bibinfo {author} {\bibfnamefont {C.}~\bibnamefont {Robert}},
  \bibinfo {author} {\bibfnamefont {M.~M.}\ \bibnamefont {Glazov}}, \bibinfo
  {author} {\bibfnamefont {F.}~\bibnamefont {Cadiz}}, \bibinfo {author}
  {\bibfnamefont {E.}~\bibnamefont {Courtade}}, \bibinfo {author}
  {\bibfnamefont {T.}~\bibnamefont {Amand}}, \bibinfo {author} {\bibfnamefont
  {D.}~\bibnamefont {Lagarde}}, \bibinfo {author} {\bibfnamefont
  {T.}~\bibnamefont {Taniguchi}}, \bibinfo {author} {\bibfnamefont
  {K.}~\bibnamefont {Watanabe}}, \bibinfo {author} {\bibfnamefont
  {B.}~\bibnamefont {Urbaszek}},\ and\ \bibinfo {author} {\bibfnamefont
  {X.}~\bibnamefont {Marie}},\ }\href
  {https://doi.org/10.1103/PhysRevLett.119.047401} {\bibfield  {journal}
  {\bibinfo  {journal} {Phys. Rev. Lett.}\ }\textbf {\bibinfo {volume} {119}},\
  \bibinfo {pages} {047401} (\bibinfo {year} {2017})}\BibitemShut {NoStop}%
\bibitem [{\citenamefont {Zhang}\ \emph {et~al.}(2017)\citenamefont {Zhang},
  \citenamefont {Cao}, \citenamefont {Lu}, \citenamefont {Lin}, \citenamefont
  {Zhang}, \citenamefont {Wang}, \citenamefont {Li}, \citenamefont {Hone},
  \citenamefont {Robinson}, \citenamefont {Smirnov}, \citenamefont {Louie},\
  and\ \citenamefont {Heinz}}]{Zhang2017}%
  \BibitemOpen
  \bibfield  {author} {\bibinfo {author} {\bibfnamefont {X.-X.}\ \bibnamefont
  {Zhang}}, \bibinfo {author} {\bibfnamefont {T.}~\bibnamefont {Cao}}, \bibinfo
  {author} {\bibfnamefont {Z.}~\bibnamefont {Lu}}, \bibinfo {author}
  {\bibfnamefont {Y.-C.}\ \bibnamefont {Lin}}, \bibinfo {author} {\bibfnamefont
  {F.}~\bibnamefont {Zhang}}, \bibinfo {author} {\bibfnamefont
  {Y.}~\bibnamefont {Wang}}, \bibinfo {author} {\bibfnamefont {Z.}~\bibnamefont
  {Li}}, \bibinfo {author} {\bibfnamefont {J.~C.}\ \bibnamefont {Hone}},
  \bibinfo {author} {\bibfnamefont {J.~A.}\ \bibnamefont {Robinson}}, \bibinfo
  {author} {\bibfnamefont {D.}~\bibnamefont {Smirnov}}, \bibinfo {author}
  {\bibfnamefont {S.~G.}\ \bibnamefont {Louie}},\ and\ \bibinfo {author}
  {\bibfnamefont {T.~F.}\ \bibnamefont {Heinz}},\ }\href
  {https://doi.org/10.1038/nnano.2017.105} {\bibfield  {journal} {\bibinfo
  {journal} {Nature Nanotechnology}\ }\textbf {\bibinfo {volume} {12}},\
  \bibinfo {pages} {883} (\bibinfo {year} {2017})}\BibitemShut {NoStop}%
\bibitem [{\citenamefont {Lu}\ \emph {et~al.}(2019)\citenamefont {Lu},
  \citenamefont {Rhodes}, \citenamefont {Li}, \citenamefont {Tuan},
  \citenamefont {Jiang}, \citenamefont {Ludwig}, \citenamefont {Jiang},
  \citenamefont {Lian}, \citenamefont {Shi}, \citenamefont {Hone},
  \citenamefont {Dery},\ and\ \citenamefont {Smirnov}}]{Lu_2019}%
  \BibitemOpen
  \bibfield  {author} {\bibinfo {author} {\bibfnamefont {Z.}~\bibnamefont
  {Lu}}, \bibinfo {author} {\bibfnamefont {D.}~\bibnamefont {Rhodes}}, \bibinfo
  {author} {\bibfnamefont {Z.}~\bibnamefont {Li}}, \bibinfo {author}
  {\bibfnamefont {D.~V.}\ \bibnamefont {Tuan}}, \bibinfo {author}
  {\bibfnamefont {Y.}~\bibnamefont {Jiang}}, \bibinfo {author} {\bibfnamefont
  {J.}~\bibnamefont {Ludwig}}, \bibinfo {author} {\bibfnamefont
  {Z.}~\bibnamefont {Jiang}}, \bibinfo {author} {\bibfnamefont
  {Z.}~\bibnamefont {Lian}}, \bibinfo {author} {\bibfnamefont {S.-F.}\
  \bibnamefont {Shi}}, \bibinfo {author} {\bibfnamefont {J.}~\bibnamefont
  {Hone}}, \bibinfo {author} {\bibfnamefont {H.}~\bibnamefont {Dery}},\ and\
  \bibinfo {author} {\bibfnamefont {D.}~\bibnamefont {Smirnov}},\ }\href
  {https://doi.org/10.1088/2053-1583/ab5614} {\bibfield  {journal} {\bibinfo
  {journal} {2D Materials}\ }\textbf {\bibinfo {volume} {7}},\ \bibinfo {pages}
  {015017} (\bibinfo {year} {2019})}\BibitemShut {NoStop}%
\bibitem [{\citenamefont {Robert}\ \emph {et~al.}(2020)\citenamefont {Robert},
  \citenamefont {Han}, \citenamefont {Kapuscinski}, \citenamefont {Delhomme},
  \citenamefont {Faugeras}, \citenamefont {Amand}, \citenamefont {Molas},
  \citenamefont {Bartos}, \citenamefont {Watanabe}, \citenamefont {Taniguchi},
  \citenamefont {Urbaszek}, \citenamefont {Potemski},\ and\ \citenamefont
  {Marie}}]{Robert2020MeasurementOT}%
  \BibitemOpen
  \bibfield  {author} {\bibinfo {author} {\bibfnamefont {C.}~\bibnamefont
  {Robert}}, \bibinfo {author} {\bibfnamefont {B.}~\bibnamefont {Han}},
  \bibinfo {author} {\bibfnamefont {P.}~\bibnamefont {Kapuscinski}}, \bibinfo
  {author} {\bibfnamefont {A.}~\bibnamefont {Delhomme}}, \bibinfo {author}
  {\bibfnamefont {C.}~\bibnamefont {Faugeras}}, \bibinfo {author}
  {\bibfnamefont {T.}~\bibnamefont {Amand}}, \bibinfo {author} {\bibfnamefont
  {M.~R.}\ \bibnamefont {Molas}}, \bibinfo {author} {\bibfnamefont
  {M.}~\bibnamefont {Bartos}}, \bibinfo {author} {\bibfnamefont
  {K.}~\bibnamefont {Watanabe}}, \bibinfo {author} {\bibfnamefont
  {T.}~\bibnamefont {Taniguchi}}, \bibinfo {author} {\bibfnamefont
  {B.}~\bibnamefont {Urbaszek}}, \bibinfo {author} {\bibfnamefont
  {M.}~\bibnamefont {Potemski}},\ and\ \bibinfo {author} {\bibfnamefont
  {X.}~\bibnamefont {Marie}},\ }\href
  {https://doi.org/10.1038/s41467-020-17608-4} {\bibfield  {journal} {\bibinfo
  {journal} {Nature Communications}\ }\textbf {\bibinfo {volume} {11}},\
  \bibinfo {pages} {4037} (\bibinfo {year} {2020})}\BibitemShut {NoStop}%
\bibitem [{\citenamefont {Zinkiewicz}\ \emph {et~al.}(2020)\citenamefont
  {Zinkiewicz}, \citenamefont {Slobodeniuk}, \citenamefont {Kazimierczuk},
  \citenamefont {Kapuściński}, \citenamefont {Oreszczuk}, \citenamefont
  {Grzeszczyk}, \citenamefont {Bartos}, \citenamefont {Nogajewski},
  \citenamefont {Watanabe}, \citenamefont {Taniguchi}, \citenamefont
  {Faugeras}, \citenamefont {Kossacki}, \citenamefont {Potemski}, \citenamefont
  {Babiński},\ and\ \citenamefont {Molas}}]{zinkiewicz2020neutral}%
  \BibitemOpen
  \bibfield  {author} {\bibinfo {author} {\bibfnamefont {M.}~\bibnamefont
  {Zinkiewicz}}, \bibinfo {author} {\bibfnamefont {A.~O.}\ \bibnamefont
  {Slobodeniuk}}, \bibinfo {author} {\bibfnamefont {T.}~\bibnamefont
  {Kazimierczuk}}, \bibinfo {author} {\bibfnamefont {P.}~\bibnamefont
  {Kapuściński}}, \bibinfo {author} {\bibfnamefont {K.}~\bibnamefont
  {Oreszczuk}}, \bibinfo {author} {\bibfnamefont {M.}~\bibnamefont
  {Grzeszczyk}}, \bibinfo {author} {\bibfnamefont {M.}~\bibnamefont {Bartos}},
  \bibinfo {author} {\bibfnamefont {K.}~\bibnamefont {Nogajewski}}, \bibinfo
  {author} {\bibfnamefont {K.}~\bibnamefont {Watanabe}}, \bibinfo {author}
  {\bibfnamefont {T.}~\bibnamefont {Taniguchi}}, \bibinfo {author}
  {\bibfnamefont {C.}~\bibnamefont {Faugeras}}, \bibinfo {author}
  {\bibfnamefont {P.}~\bibnamefont {Kossacki}}, \bibinfo {author}
  {\bibfnamefont {M.}~\bibnamefont {Potemski}}, \bibinfo {author}
  {\bibfnamefont {A.}~\bibnamefont {Babiński}},\ and\ \bibinfo {author}
  {\bibfnamefont {M.~R.}\ \bibnamefont {Molas}},\ }\href
  {https://doi.org/10.1039/D0NR04243A} {\bibfield  {journal} {\bibinfo
  {journal} {Nanoscale}\ }\textbf {\bibinfo {volume} {12}},\ \bibinfo {pages}
  {18153} (\bibinfo {year} {2020})}\BibitemShut {NoStop}%
\bibitem [{\citenamefont {Bayer}\ \emph {et~al.}(2002)\citenamefont {Bayer},
  \citenamefont {Ortner}, \citenamefont {Stern}, \citenamefont {Kuther},
  \citenamefont {Gorbunov}, \citenamefont {Forchel}, \citenamefont {Hawrylak},
  \citenamefont {Fafard}, \citenamefont {Hinzer}, \citenamefont {Reinecke},
  \citenamefont {Walck}, \citenamefont {Reithmaier}, \citenamefont {Klopf},\
  and\ \citenamefont {Sch\"afer}}]{Bayer2002}%
  \BibitemOpen
  \bibfield  {author} {\bibinfo {author} {\bibfnamefont {M.}~\bibnamefont
  {Bayer}}, \bibinfo {author} {\bibfnamefont {G.}~\bibnamefont {Ortner}},
  \bibinfo {author} {\bibfnamefont {O.}~\bibnamefont {Stern}}, \bibinfo
  {author} {\bibfnamefont {A.}~\bibnamefont {Kuther}}, \bibinfo {author}
  {\bibfnamefont {A.~A.}\ \bibnamefont {Gorbunov}}, \bibinfo {author}
  {\bibfnamefont {A.}~\bibnamefont {Forchel}}, \bibinfo {author} {\bibfnamefont
  {P.}~\bibnamefont {Hawrylak}}, \bibinfo {author} {\bibfnamefont
  {S.}~\bibnamefont {Fafard}}, \bibinfo {author} {\bibfnamefont
  {K.}~\bibnamefont {Hinzer}}, \bibinfo {author} {\bibfnamefont {T.~L.}\
  \bibnamefont {Reinecke}}, \bibinfo {author} {\bibfnamefont {S.~N.}\
  \bibnamefont {Walck}}, \bibinfo {author} {\bibfnamefont {J.~P.}\ \bibnamefont
  {Reithmaier}}, \bibinfo {author} {\bibfnamefont {F.}~\bibnamefont {Klopf}},\
  and\ \bibinfo {author} {\bibfnamefont {F.}~\bibnamefont {Sch\"afer}},\ }\href
  {https://doi.org/10.1103/PhysRevB.65.195315} {\bibfield  {journal} {\bibinfo
  {journal} {Phys. Rev. B}\ }\textbf {\bibinfo {volume} {65}},\ \bibinfo
  {pages} {195315} (\bibinfo {year} {2002})}\BibitemShut {NoStop}%
\bibitem [{\citenamefont {Kowalik}\ \emph {et~al.}(2007)\citenamefont
  {Kowalik}, \citenamefont {Krebs}, \citenamefont {Golnik}, \citenamefont
  {Suffczy\ifmmode~\acute{n}\else \'{n}\fi{}ski}, \citenamefont {Wojnar},
  \citenamefont {Kossut}, \citenamefont {Gaj},\ and\ \citenamefont
  {Voisin}}]{Kowalik2007}%
  \BibitemOpen
  \bibfield  {author} {\bibinfo {author} {\bibfnamefont {K.}~\bibnamefont
  {Kowalik}}, \bibinfo {author} {\bibfnamefont {O.}~\bibnamefont {Krebs}},
  \bibinfo {author} {\bibfnamefont {A.}~\bibnamefont {Golnik}}, \bibinfo
  {author} {\bibfnamefont {J.}~\bibnamefont {Suffczy\ifmmode~\acute{n}\else
  \'{n}\fi{}ski}}, \bibinfo {author} {\bibfnamefont {P.}~\bibnamefont
  {Wojnar}}, \bibinfo {author} {\bibfnamefont {J.}~\bibnamefont {Kossut}},
  \bibinfo {author} {\bibfnamefont {J.~A.}\ \bibnamefont {Gaj}},\ and\ \bibinfo
  {author} {\bibfnamefont {P.}~\bibnamefont {Voisin}},\ }\href
  {https://doi.org/10.1103/PhysRevB.75.195340} {\bibfield  {journal} {\bibinfo
  {journal} {Phys. Rev. B}\ }\textbf {\bibinfo {volume} {75}},\ \bibinfo
  {pages} {195340} (\bibinfo {year} {2007})}\BibitemShut {NoStop}%
\bibitem [{\citenamefont {Seguin}\ \emph {et~al.}(2005)\citenamefont {Seguin},
  \citenamefont {Schliwa}, \citenamefont {Rodt}, \citenamefont {P\"otschke},
  \citenamefont {Pohl},\ and\ \citenamefont {Bimberg}}]{Seguin2005}%
  \BibitemOpen
  \bibfield  {author} {\bibinfo {author} {\bibfnamefont {R.}~\bibnamefont
  {Seguin}}, \bibinfo {author} {\bibfnamefont {A.}~\bibnamefont {Schliwa}},
  \bibinfo {author} {\bibfnamefont {S.}~\bibnamefont {Rodt}}, \bibinfo {author}
  {\bibfnamefont {K.}~\bibnamefont {P\"otschke}}, \bibinfo {author}
  {\bibfnamefont {U.~W.}\ \bibnamefont {Pohl}},\ and\ \bibinfo {author}
  {\bibfnamefont {D.}~\bibnamefont {Bimberg}},\ }\href
  {https://doi.org/10.1103/PhysRevLett.95.257402} {\bibfield  {journal}
  {\bibinfo  {journal} {Phys. Rev. Lett.}\ }\textbf {\bibinfo {volume} {95}},\
  \bibinfo {pages} {257402} (\bibinfo {year} {2005})}\BibitemShut {NoStop}%
\bibitem [{\citenamefont {Srivastava}\ \emph {et~al.}(2008)\citenamefont
  {Srivastava}, \citenamefont {Htoon}, \citenamefont {Klimov},\ and\
  \citenamefont {Kono}}]{Srivastava2008}%
  \BibitemOpen
  \bibfield  {author} {\bibinfo {author} {\bibfnamefont {A.}~\bibnamefont
  {Srivastava}}, \bibinfo {author} {\bibfnamefont {H.}~\bibnamefont {Htoon}},
  \bibinfo {author} {\bibfnamefont {V.~I.}\ \bibnamefont {Klimov}},\ and\
  \bibinfo {author} {\bibfnamefont {J.}~\bibnamefont {Kono}},\ }\href
  {https://doi.org/10.1103/PhysRevLett.101.087402} {\bibfield  {journal}
  {\bibinfo  {journal} {Phys. Rev. Lett.}\ }\textbf {\bibinfo {volume} {101}},\
  \bibinfo {pages} {087402} (\bibinfo {year} {2008})}\BibitemShut {NoStop}%
\bibitem [{\citenamefont {Chenet}\ \emph {et~al.}(2015)\citenamefont {Chenet},
  \citenamefont {Aslan}, \citenamefont {Huang}, \citenamefont {Fan},
  \citenamefont {van~der Zande}, \citenamefont {Heinz},\ and\ \citenamefont
  {Hone}}]{Chenet2015}%
  \BibitemOpen
  \bibfield  {author} {\bibinfo {author} {\bibfnamefont {D.~A.}\ \bibnamefont
  {Chenet}}, \bibinfo {author} {\bibfnamefont {O.~B.}\ \bibnamefont {Aslan}},
  \bibinfo {author} {\bibfnamefont {P.~Y.}\ \bibnamefont {Huang}}, \bibinfo
  {author} {\bibfnamefont {C.}~\bibnamefont {Fan}}, \bibinfo {author}
  {\bibfnamefont {A.~M.}\ \bibnamefont {van~der Zande}}, \bibinfo {author}
  {\bibfnamefont {T.~F.}\ \bibnamefont {Heinz}},\ and\ \bibinfo {author}
  {\bibfnamefont {J.~C.}\ \bibnamefont {Hone}},\ }\href
  {https://doi.org/10.1021/acs.nanolett.5b00910} {\bibfield  {journal}
  {\bibinfo  {journal} {Nano Letters}\ }\textbf {\bibinfo {volume} {15}},\
  \bibinfo {pages} {5667} (\bibinfo {year} {2015})}\BibitemShut {NoStop}%
\bibitem [{\citenamefont {Liang}\ \emph {et~al.}(2009)\citenamefont {Liang},
  \citenamefont {Chan}, \citenamefont {Tiong}, \citenamefont {Huang},
  \citenamefont {Chen}, \citenamefont {Dumcenco},\ and\ \citenamefont
  {Ho}}]{LIANG200994}%
  \BibitemOpen
  \bibfield  {author} {\bibinfo {author} {\bibfnamefont {C.}~\bibnamefont
  {Liang}}, \bibinfo {author} {\bibfnamefont {Y.}~\bibnamefont {Chan}},
  \bibinfo {author} {\bibfnamefont {K.}~\bibnamefont {Tiong}}, \bibinfo
  {author} {\bibfnamefont {Y.}~\bibnamefont {Huang}}, \bibinfo {author}
  {\bibfnamefont {Y.}~\bibnamefont {Chen}}, \bibinfo {author} {\bibfnamefont
  {D.}~\bibnamefont {Dumcenco}},\ and\ \bibinfo {author} {\bibfnamefont
  {C.}~\bibnamefont {Ho}},\ }\href
  {https://doi.org/https://doi.org/10.1016/j.jallcom.2008.09.175} {\bibfield
  {journal} {\bibinfo  {journal} {Journal of Alloys and Compounds}\ }\textbf
  {\bibinfo {volume} {480}},\ \bibinfo {pages} {94} (\bibinfo {year} {2009})},\
  \bibinfo {note} {proceedings of the 16th International Conference on Solid
  Compounds of Transition Elements (SCTE 2008)}\BibitemShut {NoStop}%
\end{thebibliography}%

\end{document}